\pdfoutput=1
\documentclass[aps,prd,amsmath,floats,floatfix, twocolumn,
superscriptaddress,nofootinbib,showpacs,longbibliography]{revtex4-1}
\usepackage[T1]{fontenc}
\usepackage[utf8]{inputenc}
\usepackage{txfonts}
\usepackage{verbatim}
\usepackage[dvipsnames, usenames]{xcolor}
\definecolor{linkcolor}{rgb}{0.0,0.3,0.5}
\usepackage[hypertexnames=false, unicode, colorlinks=true, linkcolor=linkcolor,
citecolor=linkcolor, filecolor=linkcolor,urlcolor=linkcolor,
pdfusetitle]{hyperref}
\usepackage[all]{hypcap}
\usepackage{graphicx}
\usepackage{xspace}
\usepackage{amssymb}
\usepackage{microtype}
\usepackage{subfigure}

\usepackage{url}

\usepackage[english]{babel}
\usepackage{blindtext}

\usepackage[normalem]{ulem} %for \sout
\usepackage{bm} % boldmath
\usepackage{mathrsfs}
\usepackage{xspace} % for \xspace

\DeclareMathAlphabet{\mathpzc}{OT1}{pzc}{m}{it}
\usepackage[nomessages]{fp}

\newcommand{\sk}[1]{}

\begin{document}
\title{Data-driven extraction, phenomenology and modeling \\of eccentric harmonics in binary black hole merger waveforms}
\newcommand{\KITP}{\affiliation{Kavli Institute for Theoretical Physics, University of California Santa Barbara, Kohn Hall, Lagoon Rd, Santa Barbara, CA 93106}} 
\newcommand{\TAPIR}{\affiliation{Theoretical AstroPhysics Including Relativity and Cosmology, California Institute of Technology, Pasadena, California, USA}}
\author{Tousif Islam}
\email{tousifislam@ucsb.edu}
\KITP
\TAPIR
\author{Tejaswi Venumadhav}
\affiliation{\mbox{Department of Physics, University of California at Santa Barbara, Santa Barbara, CA 93106, USA}}
\affiliation{\mbox{International Centre for Theoretical Sciences, Tata Institute of Fundamental Research, Bangalore 560089, India}}
\author{Ajit Kumar Mehta}
\affiliation{\mbox{Department of Physics, University of California at Santa Barbara, Santa Barbara, CA 93106, USA}}
\author{Isha Anantpurkar}
\affiliation{\mbox{Department of Physics, University of California at Santa Barbara, Santa Barbara, CA 93106, USA}}
\author{Digvijay Wadekar}
\affiliation{\mbox{Department of Physics and Astronomy, Johns Hopkins University,
3400 N. Charles Street, Baltimore, Maryland, 21218, USA}}
\affiliation{\mbox{School of Natural Sciences, Institute for Advanced Study, 1 Einstein Drive, Princeton, NJ 08540, USA}}
\author{Javier Roulet}
\affiliation{\mbox{TAPIR, Walter Burke Institute for Theoretical Physics, California Institute of Technology, Pasadena, CA 91125, USA}}
\author{Jonathan Mushkin}
\affiliation{\mbox{Department of Particle Physics \& Astrophysics, Weizmann Institute of Science, Rehovot 76100, Israel}}
\author{Barak Zackay}
\affiliation{\mbox{Department of Particle Physics \& Astrophysics, Weizmann Institute of Science, Rehovot 76100, Israel}}
\author{Matias Zaldarriaga}
\affiliation{\mbox{School of Natural Sciences, Institute for Advanced Study, 1 Einstein Drive, Princeton, NJ 08540, USA}}

% Because hyperref only gets the *last* author, we need to be explicit.
\hypersetup{pdfauthor={Islam et al.}}

\date{\today}

%==========================================================================
%==========================================================================
\begin{abstract}
Newtonian and post-Newtonian (PN) calculations suggest that each spherical harmonic mode of the gravitational waveforms (radiation) emitted by eccentric binaries can be further decomposed into several eccentricity-induced modes (indexed by $j=1$ to $j=\infty$), referred to as eccentric harmonics. These harmonics exhibit monotonically time-varying amplitudes and instantaneous frequencies, unlike the full eccentric spherical harmonic modes. 
However, computing or extracting these harmonics are not straightforward in current numerical relativity (NR) simulations and eccentric waveform models. 
To address this, ~\citet{Patterson:2024vbo} have developed a framework to extract the eccentric harmonics directly from effective-one-body formalism waveforms. In this paper, we build on the ideas presented in ~\citet{Patterson:2024vbo} and propose a data-driven framework, utilizing singular-value decomposition (SVD), that incorporates additional features based on PN intuition to ensure monotonicity in the extracted harmonics. 
We further demonstrate that the phase (frequency) of these harmonics is simply $j\phi_{\lambda}+\phi_{\rm ecc}$ ($jf_{\lambda}+f_{\rm ecc}$) where $\phi_{\lambda}$ ($f_{\lambda}$) is related to the secular orbital phase (frequency) and $\phi_{\rm ecc}$ ($f_{\rm ecc}$) is an additional phase (frequency) that only depends on the eccentricity. We also provide simple analytical fits to obtain the harmonics as a function of the mean anomaly.
These relations may prove useful in constructing faithful models (such as Ref.~\cite{Islam2025InPrep}) that can be employed in cheap and efficient searches and parameter estimation of eccentric mergers. Our framework is modular and can be extended for any other eccentric waveform models or simulation frameworks. The framework is available through the \texttt{gwMiner} package~\footnote{\href{https://github.com/tousifislam/gwMiner}{https://github.com/tousifislam/gwMiner} (available on request)}.
\end{abstract}
%==========================================================================
%==========================================================================
\maketitle

%==========================================================================
%==========================================================================
%==========================================================================
\section{Introduction}
Understanding the phenomenology of eccentric binary black hole mergers and modeling them is one of the most active topics of interest in the gravitational wave community. This is partly because the strain data obtained by the LIGO-Virgo-KAGRA Collaboration has revealed around 100 binary black holes (BBH) mergers~\cite{Harry:2010zz,VIRGO:2014yos,KAGRA:2020tym,LIGOScientific:2018mvr,LIGOScientific:2020ibl,LIGOScientific:2021usb,LIGOScientific:2021djp}, but none of those have been conclusively identified as eccentric. Possible eccentric interpretation of some detected BBHs is however investigated in Refs.~\cite{Romero-Shaw:2020thy,Gayathri:2020coq,Gamba:2021gap,Ramos-Buades:2023yhy,Gupte:2024jfe,Morras:2025xfu}.
Their rarity has motivated researchers to be prepared for their detection in future gravitational wave observations. Eccentric binaries are particularly interesting as they are expected to be more common in dense galaxies and globular clusters, carrying significant information about these environments through gravitational waves. At the population level, the eccentricity distribution of binaries is also expected to vary depending on their formation channels~\cite{Rodriguez:2017pec,Rodriguez:2018pss,Samsing:2017xmd,Zevin:2018kzq,Zevin:2021rtf,Samsing:2020tda}.

In recent times, significant multidimensional efforts are underway to advance our understanding of eccentric binaries. For example, an increasing number of eccentric numerical relativity (NR) simulations are being performed and published in the public domain~\cite{Mroue:2010re, Healy:2017zqj,Buonanno:2006ui,Husa:2007rh,Ramos-Buades:2018azo,Ramos-Buades:2019uvh,Purrer:2012wy,Bonino:2024xrv,Ramos-Buades:2022lgf}. Standard waveform models such as effective-one-body (EOB) models and NR surrogates have started incorporating eccentricity using many of these simulations~\cite{Tiwari:2019jtz, Huerta:2014eca, Moore:2016qxz, Damour:2004bz, Konigsdorffer:2006zt, Memmesheimer:2004cv,Hinder:2017sxy, Cho:2021oai,Chattaraj:2022tay,Hinderer:2017jcs,Cao:2017ndf,Chiaramello:2020ehz,Albanesi:2023bgi,Albanesi:2022xge,Riemenschneider:2021ppj,Chiaramello:2020ehz,Ramos-Buades:2021adz,Liu:2023ldr,Huerta:2016rwp,Huerta:2017kez,Joshi:2022ocr,Setyawati:2021gom,Wang:2023ueg,Islam:2021mha,Carullo:2023kvj,Nagar:2021gss,Tanay:2016zog,Paul:2024ujx,Manna:2024ycx,Gamboa:2024hli,Planas:2025feq}, often aided by new theoretical calculations within the post-Newtonian (PN) formalism~\cite{Arun:2009mc,Tanay:2016zog,Paul:2022xfy,Henry:2023tka,Gamboa:2024imd}. On the phenomenology side, new insights are being obtained regarding the effect of eccentricity on different radiation quantities. Furthermore, efforts are also underway to standardize the definition of eccentricity~\cite{blanchet:2013haa,Mroue:2010re, Healy:2017zqj,Mora:2002gf,Ramos-Buades:2021adz,Islam:2021mha,Ramos-Buades:2019uvh,Shaikh:2023ypz,Knee:2022hth,Boschini:2024scu}.

One of the challenges in modeling eccentricity for BBHs is that eccentricity introduces oscillatory features in waveform quantities, and these additional complex features are typically hard to model. While several strategies have been discussed on how to efficiently decompose the oscillatory features into much simpler components and then model them~\cite{Islam:2021mha,Islam:2024rhm}, the problem remains reasonably complex in nature. 

In pursuit of such simpler features, we note that Newtonian calculations suggest that different (spherical harmonic) modes of gravitational waveforms from eccentric binaries can be written as a superposition of several smooth harmonics, called eccentric harmonics~\cite{Yunes:2009yz}. If these modes were readily available either from NR or EOB calculations, modeling radiation from eccentric binaries would become significantly simpler. However, the challenge is that a clear calculation of these harmonics is not available even within PN frameworks~\cite{VanDenBroeck:2006qu,Arun:2007qv,Seto:2001pg,2012ApJ74537V}. Refs.~\cite{Hegde:2023yoz,Bose:2021pcw} provided phenomenological scaling relations for the eccentric harmonics' time-frequency track based on a post-Newtonian waveform model named \texttt{EccentricTD}.
Even if we can identify these harmonics in existing PN expressions, we would still need to extend these calculations to merger and ringdown, which is a difficult task. 

\begin{figure*}
\includegraphics[width=\textwidth]{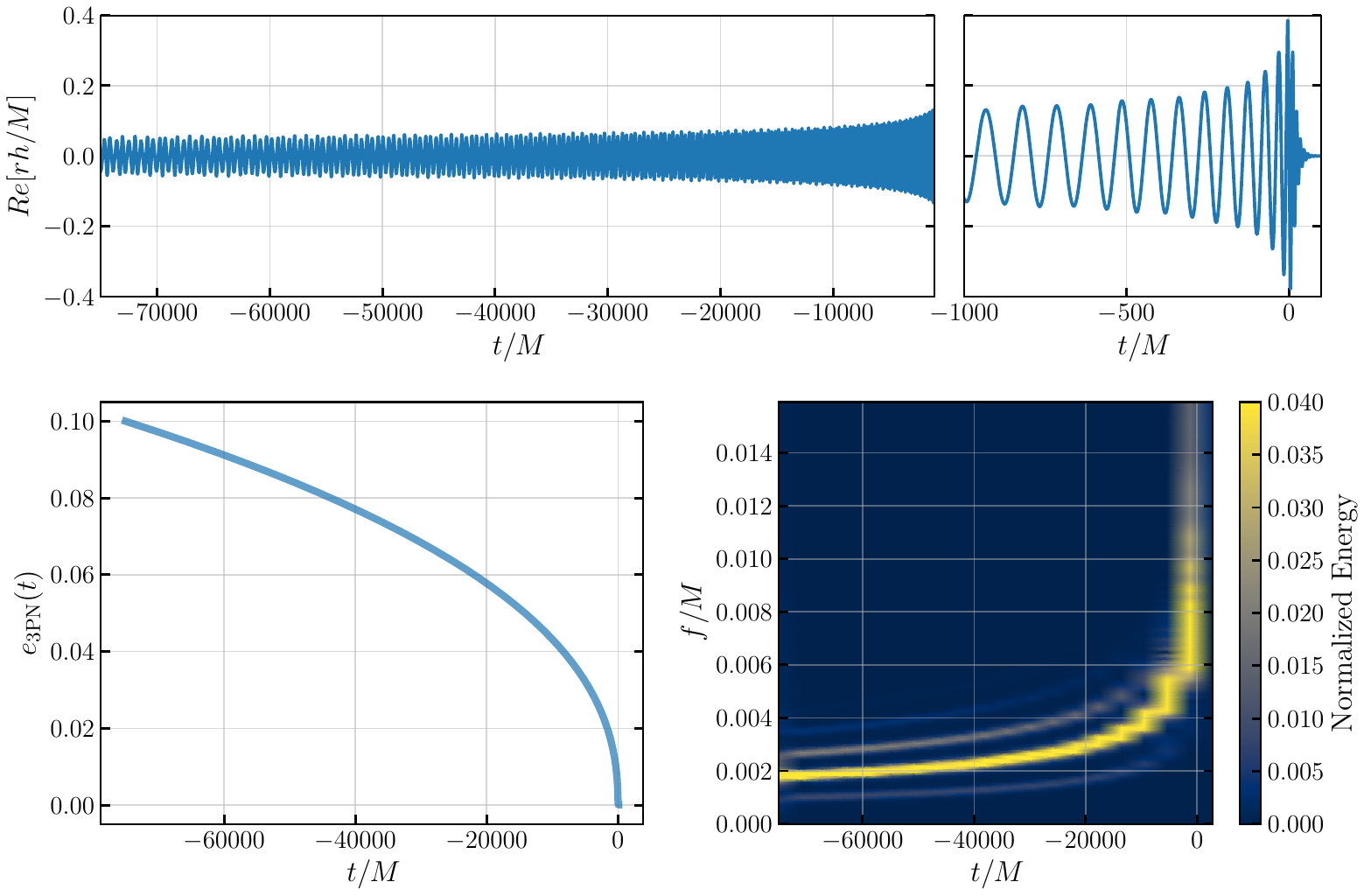}
\caption{\textit{Upper panel:} We show the real part of the quadrupolar mode gravitational waveform generated from a non-spinning binary characterized by $q=1$, $e_{\rm ref}=0.1$, and $l_{\rm ref}=\pi$ using \texttt{TEOBResums} model~\cite{Chiaramello:2020ehz}. The reference time is chosen to be $t = -75000M$ while $t=0$ denotes the time of merger. Furthermore, we present the eccentricity evolution of the binary as a function of time in the \textit{lower left panel} and a time-frequency spectrogram obtained from the STFT in the \textit{lower right panel}. Details are in Section~\ref{sec:filter_method}.}
\label{fig:spectogram}
\end{figure*}

Recently, Ref.~\cite{Patterson:2024vbo} has presented a framework to compute these eccentric harmonics from EOB waveforms by combining a set of waveforms with the same eccentricities but different phase rotations.
In this paper, we utilize insights from Newtonian and PN calculations to develop two distinct frameworks, that further refines and extends the methods presented in Ref.~\cite{Patterson:2024vbo}, for extracting eccentric harmonics in gravitational waveforms. While our current framework is demonstrated using EOB waveforms, it can be generalized to other types of waveforms as well. We provide the details of our frameworks in Section~\ref{sec:extraction}. Section~\ref{sec:filter_method} focuses on building a framework that utilizes Newtonian expectations of the eccentric harmonic frequencies and develops low-pass filters to isolate those harmonics individually for any given spherical harmonic mode. Section~\ref{sec:svd_method} presents a complementary method that uses Singular Value Decomposition (SVD) to extract the harmonics from an ensemble of eccentric waveforms. Then, in Section~\ref{sec:filter_vs_svd}, we compare the output of these two methods one-by-one. In Section~\ref{sec:phenomenology}, we delve into the phenomenology of the eccentric harmonics. We compare the phase of the dominant eccentric harmonic $j=2$ with the circular waveform and highlight key differences. We then analyze the hierarchy of the amplitudes of different eccentric harmonics and discover a simple phase and frequency relation between these harmonics. Finally, we provide a simple but effective modeling strategy to include the effect of the mean anomaly parameter in eccentric waveform models in Section~\ref{sec:model_mean_anomaly}. We discuss potential implications and future directions in Section~\ref{sec:discussion}.

%==========================================================================
%==========================================================================
%==========================================================================
\section{Extraction of eccentric harmonics}
\label{sec:extraction}
The gravitational radiation from a binary black hole (BBH) merger is expressed as a complex waveform $h(t)$, constructed from the two orthogonal polarizations:
\begin{align}
h(t, \theta, \phi; \boldsymbol{\lambda}) &= h_{+}(t, \theta, \phi; \boldsymbol{\lambda}) - i h_{\times}(t, \theta, \phi; \boldsymbol{\lambda}),
\end{align}
where $t$ represents time, $\theta$ and $\phi$ are angles on the merger's sky, and $\boldsymbol{\lambda}$ is the set of intrinsic parameters describing the binary. Throughout this paper, we use geometric units, measuring time in units of the total mass of the binary and setting $G = c = 1$ unless stated otherwise. The parameter set $\boldsymbol{\lambda}$ is given by:
\[\boldsymbol{\lambda} := \{q, \boldsymbol{\chi}_1, \boldsymbol{\chi}_2, e_{\rm ref}, l_{\rm ref} \},\]
where $q := m_1 / m_2$ is the mass ratio, with $m_1$ and $m_2$ denoting the masses of the larger and smaller black holes, respectively. The vectors $\boldsymbol{\chi}_1$ and $\boldsymbol{\chi}_2$ represent the three-dimensional spin components of the two black holes. Eccentricity is characterized by two parameters: $e_{\rm ref}$, the eccentricity, and $l_{\rm ref}$, the mean anomaly, both defined at a chosen reference time or frequency. While multiple definitions of eccentricity exist~\cite{Mroue:2010re, Healy:2017zqj, Shaikh:2023ypz, Knee:2022hth, 2023PhRvD.108l4063R}, any consistent choice applied uniformly across all relevant binaries will be adequate for practical purposes.

The full complex waveform $h(t)$ consists of several $-2$ spin-weighted complex-valued spherical harmonic modes indexed by $(\ell, m)$~\cite{Maggiore:2007ulw, Maggiore:2018sht}:  
\begin{align}
h(t, \theta, \phi; \boldsymbol{\lambda}) &= \sum_{\ell=2}^\infty \sum_{m=-\ell}^{\ell} h_{\ell m}(t; \boldsymbol\lambda)  _{-2}Y_{\ell m}(\theta,\phi).
\label{hmodes}
\end{align}  
Each harmonic is typically referred to as a mode. In most binaries, the dominant contribution to the radiation comes from the $\ell=2$, $m=2$ mode, commonly known as the quadrupolar mode, while all other modes are collectively referred to as higher-order modes. Each spherical harmonic mode has a real-valued amplitude $A_{\ell m}(t)$ and phase $\phi_{\ell m}(t)$ such that:
\begin{equation}
h_{\ell m}(t;\boldsymbol{\lambda}) = A_{\ell m}(t) e^{i \phi_{\ell m}(t)}.
\label{eq:amp_phase}
\end{equation}
The instantaneous angular frequency of the spherical harmonic mode then becomes $\omega_{\ell m}(t;\boldsymbol{\lambda}) = {d\phi_{\ell m}(t)}/{dt}$. Similarly, the GW frequency of any mode is $f_{\ell m}(t) = {\omega_{\ell m}(t)}/{\pi}$. 

To generate eccentric non-spinning waveforms, we use a publicly available, NR-validated EOB waveform approximant named \texttt{TEOBResumS}. Throughout this paper, we use the eccentricity definition used in \texttt{TEOBResumS} model. Details of this waveform model are accessible in Ref.~\cite{Chiaramello:2020ehz}. We access the approximant from the \texttt{eccentric} branch of \href{https://bitbucket.org/teobresums/teobresums}{https://bitbucket.org/teobresums/teobresums/} package. We align the raw EOB waveforms such that the peak amplitude of the $(2,2)$ mode occurs at $t=0$, and the phase is zero at the waveform start time.

Now, for eccentric binaries, each of these spherical harmonic modes is thought to be further decomposable into smooth harmonics that depend only on eccentricity. Their structure is expected to follow:
\begin{align}
h_{\ell m}(t; \boldsymbol\lambda) &= \sum_{j=1}^\infty h_{\ell m, j}(t; \boldsymbol\lambda),
\label{hmodes}
\end{align}
where the $j=2$ eccentric harmonic remains dominant and is closest to the circular expectation. Harmonics associated with $j \neq 2$ are excited only when the binary has residual eccentricity and are expected to diminish as eccentricity approaches zero, according to Newtonian and PN calculations. 

Once we obtain the eccentric harmonics, we calculate the amplitude, phase, instantaneous frequency and GW frequency as:
\begin{align}
A_{22,j}(t) &= |h_{22,j}|, \\
\phi_{22,j}(t) &= {\rm Arg}[h_{22,j}],\\
\omega_{22,j}(t) &= \frac{d\phi_{22,j}}{dt},\\
f_{22,j}(t) &= \frac{\omega_{22,j}}{\pi}.
\end{align}

We note that existing literature on Newtonian and PN calculations of radiation from eccentric binaries suggests that the phase of the eccentric harmonics is roughly an integer multiple of the orbital phase $\phi_{\rm orb}(t)$:
\begin{equation}
\phi_{\ell m,j}(t) \approx j \,\phi_{\rm orb}(t). 
\label{eq:phase_j_relation}
\end{equation}
Their instantaneous and GW frequency also follow\sout{s} a similar relation:
\begin{equation}
\begin{split}
\omega_{\ell m,j}(t) &\approx j\, \omega_{\rm orb}(t),\\
f_{\ell m,j}(t) &\approx j\, f_{\rm orb}(t). 
\end{split}
\label{eq:freq_j_relation}
\end{equation}
In the relativistic scenario, we expect slight deviations from these relations. However, the above relation still provides an approximate picture of the mode hierarchy. 

We demonstrate this hierarchy in Fig.~\ref{fig:spectogram}, where we show the time-frequency spectrogram (obtained from the short-time Fourier transform (STFT)) of the quadrupolar mode from an eccentric, non-spinning binary characterized by $q=1$, $e_{\rm ref}=0.1$ and $l_{\rm ref}=\pi$. The reference time is chosen to be $t = -70000M$, with $t=0$ denoting the time of merger, i.e., the time associated with the peak amplitude of the quadrupolar mode. To obtain the STFT, we use \texttt{scipy.signal.stft} module. 
We show that the spectrogram reveals at least four visibly prominent time-frequency tracks corresponding to the eccentric harmonics. The frequency evolution of these harmonics closely follows the relations shown above, with some deviations, of course.

We will use the fundamental calculations in the Newtonian (and PN) regime as a guide to extract the eccentric harmonics for eccentric binaries using fully relativistic waveforms. For simplicity, we will restrict ourselves to the quadrupolar mode in non-spinning eccentric binaries only. So, for the rest of the paper, $\boldsymbol{\chi_1}=\boldsymbol{\chi_2}=[0,0,0]$. 

%==========================================================================
%==========================================================================
\subsection{Extracting eccentric harmonics using Singular Value Decomposition}
\label{sec:svd_method}
First, we modify and extend the framework presented in Ref.~\cite{Patterson:2024vbo} for extracting these eccentric harmonics using an ensemble of waveforms characterized by the same eccentricity but with different mean anomaly values, while keeping all other parameters unchanged. 
We incorporate a set of steps to ensure that the extracted harmonics truly correspond to smooth eccentric harmonics. 
In the following texts, we will present our framework and point out to the differences and similarities with the framework presented in Ref.~\cite{Patterson:2024vbo}. 
We leave a detailed comparison of our approach to that of Ref.~\cite{Patterson:2024vbo} to Appendix~\ref{sec:patterson}.
Below, we discuss the components of our framework one by one.

%==========================================================================
\subsubsection{Calculating eccentricity evolution tracks}
To compute the eccentricity evolution of a binary as a function of the GW frequency (or equivalently, the orbital frequency), we use a 3PN approximation valid for low eccentricities ($e \leq 0.2$) and non-spinning binaries. This eccentricity evolution reads~\cite{Moore:2016qxz}:
\begin{equation}
e_{\rm 3PN}(t)=e_{\rm ref}\left(\frac{f_{\rm gw, \rm ref}}{f_{\rm gw}}\right)^{19 / 18} \frac{g\left(f_{\rm gw}\right)}{g\left(f_{\rm gw, \rm ref}\right)}
\label{eq:ecc_evolve}
\end{equation}
\begin{widetext}
\begin{equation}
\begin{aligned}
& g\left(f_{\rm gw}\right)=\left\{1+\left(-\frac{2833}{2016}+\frac{197}{72} \eta\right) f_{\rm gw}^{2 / 3}-\frac{377}{144} \pi f_{\rm gw}+\left(\frac{77006005}{24385536}-\frac{1143767}{145152} \eta+\frac{43807}{10368} \eta^2\right) f_{\rm gw}^{4 / 3}\right. \\
& +\left(\frac{9901567}{1451520}-\frac{202589}{362880} \eta\right) \pi f_{\rm gw}^{5 / 3}+\left[-\frac{33320661414619}{386266890240}+\frac{3317}{252} \gamma_E+\frac{180721}{41472} \pi^2+\left(\frac{161339510737}{8778792960}+\frac{3977}{2304} \pi^2\right) \eta\right. \\
& \left.\left.-\frac{359037739}{20901888} \eta^2+\frac{10647791}{2239488} \eta^3-\frac{87419}{3780} \ln 2+\frac{26001}{1120} \ln 3+\frac{3317}{504} \ln \left(16 f_{\rm gw}^{2 / 3}\right)\right] f_{\rm gw}^2\right\} .
\end{aligned}
\end{equation}
\end{widetext}
Here, $f_{\rm gw, ref}$ corresponds to the initial GW frequency of the reference waveform and $\eta={q}/{(1+q)^2}$ is the symmetric mass ratio. 

\begin{figure}
\includegraphics[width=\columnwidth]{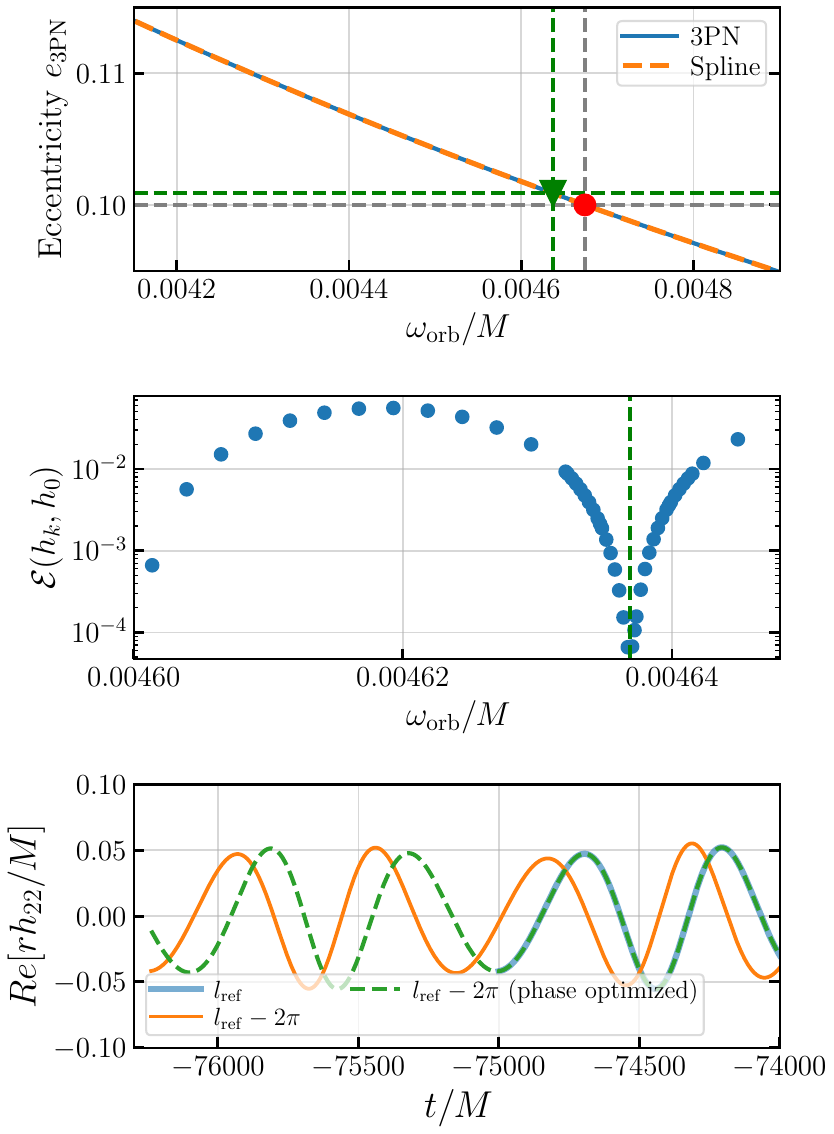}
\caption{We show the eccentricity evolution as a function of $\omega_{\rm orb}$ for $q=1$ and $e_{\rm ref}=0.1$ at $t_{\rm ref}=-75000M$. Additionally, we present the spline predictions (\textit{upper panel}) as orange dashed line. In the \textit{middle panel}, we show the errors between the reference waveform and waveforms generated with slightly smaller $\omega_{\rm orb}$ values and the associated eccentricities picked from the eccentricity track as a function of $\omega_{\rm orb}$. For $\omega_{\rm orb}=\omega_{\rm orb, left}$, this error is minimized (shown as vertical dashed line). Finally, in the lower panel, we show that the reference waveform (with $l_{\rm ref}$) and the waveform generated with initial orbital frequency of $\omega_{\rm orb, left}$ (which corresponds to a mean anomaly value of $l_{\rm ref}-2\pi$ at the reference time) matches very well after a phase rotation. Details are provided in Section~\ref{sec:same_ecc_diff_ano}.}
\label{fig:ecc_freq_evolve}
\end{figure}

%==========================================================================
\subsubsection{Generating waveforms with same eccentricity but different mean anomaly}
\label{sec:same_ecc_diff_ano}
Next, we intend to generate multiple \texttt{TEOBResumS} quadrupolar waveforms $\{h_k(t)\}$ with the same $e_{\rm ref}$ (as characterized at $t_{\rm ref}=-75000M$) but different mean anomaly values $l_{\rm ref}$ while keeping $q$, $\boldsymbol{\chi_1}$, and $\boldsymbol{\chi_2}$ fixed. However, the \texttt{TEOBResumS} model does not allow changing the mean anomaly value directly. Instead, it takes $e_{\rm ref}$ and $f_{\rm gw, ref}$ (or $\omega_{\rm orb, ref}$) as inputs and generates waveform that has initial mean anomaly fixed to $\pi$. 

To circumvent this limitation, we adopt an alternative method proposed in Ref.~\cite{Clarke:2022fma} (and later used in Ref.~\cite{Patterson:2024vbo}). Specifically, we initialize \texttt{TEOBResumS} model at a slightly smaller value of $f_{\rm gw, ref}$ (or $\omega_{\rm orb}$) with an initial eccentricity chosen using Eq.~(\ref{eq:ecc_evolve}) such that the evolved eccentricity at the reference time (or frequency) remains equal to $e_{\rm ref}$. For faster calculation, we pre-compute the frequency evolution of the eccentricity parameter for the reference binary using Eq.~(\ref{eq:ecc_evolve}) and construct a cubic spline representation to interpolate the results efficiently. Note that all these waveforms, despite having the same initial mean anomaly of $\pi$, will have different values of mean anomaly at $t_{\rm ref}$. Furthermore, if we generate a waveform that starts exactly one orbit before the reference waveform, the mean anomaly of this waveform at the reference time will be $l_{\rm ref}=-\pi$ (i.e., there will be a $2\pi$ difference with respect to the reference waveform), and both waveforms will be identical since the mean anomaly is a cyclic parameter with a periodicity of $2\pi$.

For our purposes, we first determine the initial $\omega_{\rm orb}$ (in the \texttt{TEOBResumS} approximant) corresponding to the waveform with $l_{\rm ref}=-\pi$ by generating multiple waveforms $h_k(t)$ at different starting values of $\omega_{\rm orb}$ on a coarser grid and computing the $L_2$-norm error between the reference waveform $h_0(t)$ and the generated waveforms $h_i(t)$~\cite{Blackman:2017dfb}~\footnote{To ensure that our result is not dependent on a specific detector, we do not use a detector power spectral density-weighted frequency-domain mismatch, as done in Ref.~\cite{Patterson:2024vbo}.}:
\begin{equation}
\mathcal{E}(h_0,h_k) = \frac{1}{2} \frac{\int_{t_{\rm min}}^{t_{\rm max}}|h_{0}(t) - h_{k}(t)|^2 dt}{\int_{t_{\rm min}}^{t_{\rm max}}|h_{0}(t)|^2 dt},
\label{eq:l2err}
\end{equation}
where $t_{\rm min}$ and $t_{\rm max}$ denote the initial and final times, respectively. While calculating the error, we cast the two waveforms onto the same time grid (which is the time grid of the reference waveform) and align them in phase at the start of the waveform. We then build a spline representation of the error as a function of the starting $\omega_{\rm orb}$. The spline is evaluated on a denser grid of $\omega_{\rm orb}$, and we find the value of $\omega_{\rm orb}$ that minimizes the error function. This value corresponds to a mean anomaly difference of $2\pi$. We denote this value of $\omega_{\rm orb}$ as $\omega_{\rm orb, left}$, indicating that it is associated with a waveform that starts exactly one orbit before the reference waveform.

In Fig.~\ref{fig:ecc_freq_evolve}, we show the eccentricity evolution track as a function of orbital frequency $\omega_{\rm orb}$, as calculated using a 3PN approximation for $q=1$ and $e_{\rm ref}=0.1$. We also show the faster spline representation we built for repeated faster use. Additionally, we show the errors between waveforms generated with slightly smaller $\omega_{\rm orb}$ and associated eccentricity picked from the eccentricity track as a function of $\omega_{\rm orb}$. We denote $\omega_{\rm orb, left}$ (for which the error is minimized) as a vertical line. Finally, we show that the waveform generated with an initial orbital frequency of $\omega_{\rm orb, left}$ (which corresponds to a mean anomaly value of $l_{\rm ref}-2\pi$ at the reference time) starts exactly one orbit before the reference waveform and matches well with the reference waveform.

Once we obtain $\omega_{\rm orb, left}$, we uniformly select a total of $N = 50$ values of $\omega_{\rm orb}$ within the range $[\omega_{\rm orb, left}, \omega_{\rm orb, ref}]$. We then calculate the corresponding initial eccentricities from the eccentricity-frequency track computed earlier and generate \texttt{TEOBResumS} waveforms $\{h_k(t)\}$. These 50 waveforms represent 50 different mean anomaly values at $t_{\rm ref}$ within the range $[l_{\rm ref} - 2\pi, l_{\rm ref}]$. Since these waveforms start at most one orbit before the reference waveform, the mean anomaly is expected to change approximately linearly~\cite{Clarke:2022fma,Patterson:2024vbo}. Therefore, we can reasonably assume that the mean anomaly values of these waveforms at $t_{\rm ref}$ will be uniformly distributed between $l_{\rm ref} - 2\pi$ and $l_{\rm ref}$.

%==========================================================================
\subsubsection{Performing SVD to obtain eccentric harmonic basis}
\label{sec:svd_X}
Singular value decomposition (SVD) is a widely used matrix factorization technique used in dimensionality reduction and feature extraction and has previously been used in detection and modeling of GWs. SVD decomposes any given real or complex valued matrix of size $m \times n$ into three simpler matrices~\cite{10.5555/1538674}:
\begin{equation}
X = U \Sigma V^T    
\end{equation}
where $U$ ($V$) is an $m \times m$ ($n \times n$) orthogonal matrix whose columns are called the \textit{left singular vectors} (\textit{right singular vectors}) of $X$. $V^T$ is the conjugate transpose of $V$. $\Sigma$ is an $m \times n$ diagonal matrix with non-negative real numbers (called \textit{singular values}). The singular values $\{\lambda_1,\lambda_2,..,\lambda_n\}$ in $\Sigma$ indicate the magnitude of the corresponding singular vectors $\{v_1,v_2,..,v_n\}$ (we will call this SVD basis vectors) in $V^T$. Typically both the singular values (associated SVD basis vectors) are ordered from largest to smallest (most important to least important). We can approximate any row of the $X=\{x_1,x_2,...,x_n\}$ matrix as a linear superposition of the SVD basis vectors $\{v_1,v_2,..,v_n\}$:
\begin{equation}
x_i = \sum_{i^{\prime}=1}^{i^{\prime}=n} a_{i^{\prime}} v_{i^{\prime}}
\end{equation}
where $i$ (and $i^{\prime}$) goes from $1$ to $n$. To perform the SVD, we use \texttt{scipy.linalg.svd} module~\footnote{\href{https://docs.scipy.org/doc/scipy/reference/generated/scipy.linalg.svd.html}{https://docs.scipy.org/doc/scipy/reference/generated/scipy.linalg.svd.html}}.

Once we have the carefully curated set of eccentric waveforms $\{h_k(t_k)\}$, we construct the matrix $X$, where each row corresponds to a waveform from $\{h_k(t)\}$. While constructing the $X$ matrix, we project each waveform $h_k(t_k)$ onto a common time grid that matches the time grid of the reference waveform $h_0(t_0)$~\footnote{In contrast, Ref.~\cite{Patterson:2024vbo} uses whitened waveforms to construct the 
$X$ matrix (using the PSD corresponding to the Advanced LIGO design sensitivity).}. Furthermore, we align each waveform at the beginning of this common time grid such that the initial phase is zero. 

\begin{figure}
\includegraphics[width=\columnwidth]{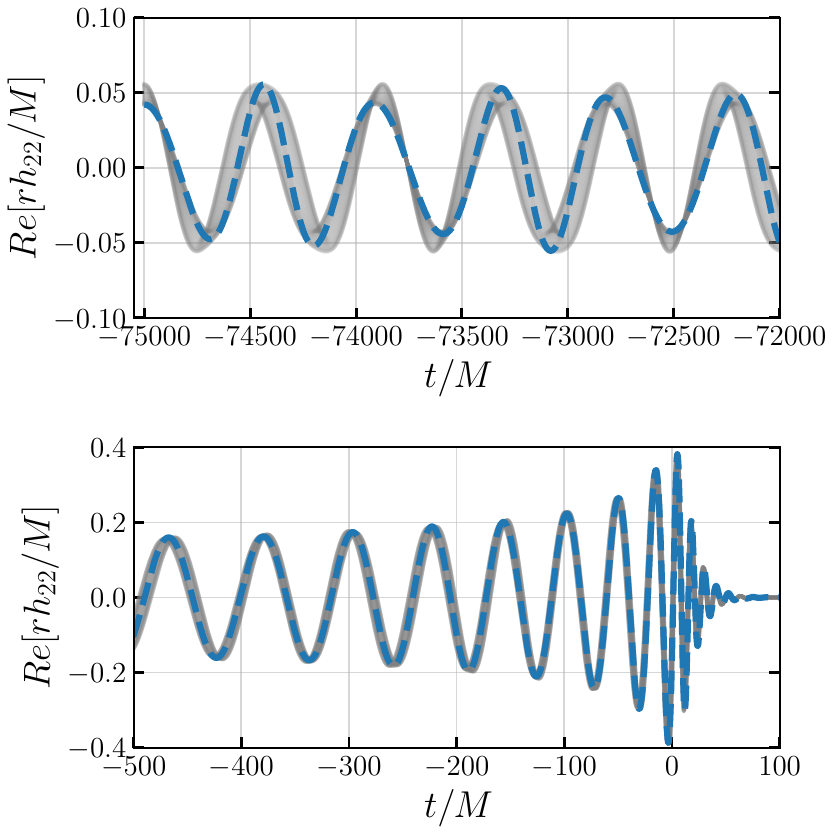}
\caption{We show the early (\textit{top panel}) and late (\textit{bottom panel}) inspiral of all 50 waveforms with \( q=1 \), \( e_{\rm ref}=0.1 \), but with different mean anomaly values \( l_{\rm ref} \). For reference, we also plot the reference waveform with \( l_{\rm ref}=\pi \). This is the set of waveforms that we pass to SVD. Details are provided in Section~\ref{sec:svd_X}. 
}
\end{figure}

We then apply SVD to extract a reduced set of basis vectors that efficiently describe binaries with the same eccentricity but different mean anomaly. The expectation is that these basis vectors will exhibit features similar to the eccentric harmonics. In other words, we hypothesize that the SVD basis vectors will correspond to the eccentric harmonic basis vectors. However, it is important to note that, in general, there is no guarantee that SVD will explicitly extract the eccentric basis, nor that the SVD basis will always have a clear physical interpretation. Nevertheless, in this case, we have constrained the SVD algorithm by ensuring that the matrix $X$ consists of waveforms with the same eccentricity. Earlier studies, such as those in Ref.~\cite{Patterson:2024vbo}, have suggested that SVD may be capable of identifying the eccentric harmonics, providing a motivation for this approach~\footnote{While Ref.~\cite{Patterson:2024vbo} draws inspiration from the SVD-based method (as we do), it does not employ SVD to obtain the harmonics due to computational complexity. Instead, Ref.~\cite{Patterson:2024vbo} attempts to use a Newtonian/PN-predicted approximations of the phases/frequencies to combine the waveforms $\{h_{i^\prime}\}$ in deriving these harmonics.}.

\begin{figure*}
\includegraphics[width=\textwidth]{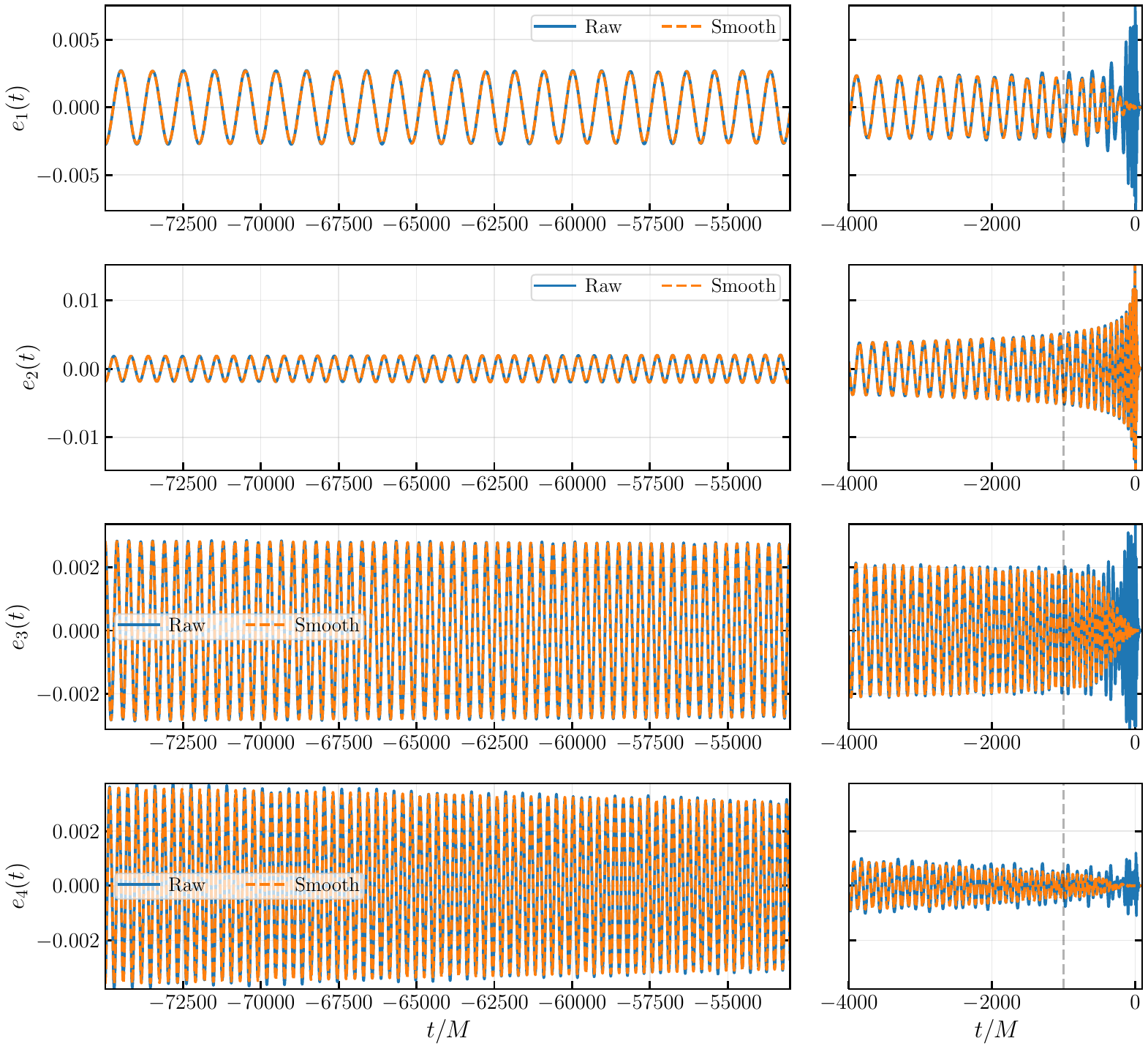}
\caption{We show the early (\textit{left panel}; for the time window of $[-75000,-52000]M$) and late (\textit{right panel}; for the time window of $[-5000,100]M$) inspiral of the four most important eccentric harmonic basis vectors $[e_1,e_2,e_3,e_4]$ for $q=1$ and $e_{\rm ref}=0.1$ binaries. While basis vectors obtained directly from SVD may exhibit a bit of noises (solid blue lines), we smooth them out (dashed orange lines). Details are provided in Section~\ref{sec:svd_X} and Section~\ref{sec:svd_smooth}. Reconstruction error after performing the smoothing is negligible (as the non-dominant eccentric harmonics goes to zero close to the merger) and is explicitly shown in Fig.~\ref{fig:SVD_recon}.}
\label{fig:q_1_ecc_0.1_SVD_basis}
\end{figure*}

As expected, we observe that the SVD basis vectors resemble interesting features that we seek in eccentric harmonics. For example, the leading basis vector looks just like the circular waveform. We identify this as the $j=2$ harmonic basis. To determine which basis corresponds to which value of $j$, we calculate their initial frequency and check its closest integer value. This helps us assign a proper value of $j$ to each harmonic basis. Going forward, we will refer to these SVD bases as $\{e_1(t), e_2(t), .., e_j(t), ..\}$ where $j$ is the eccentric harmonic index. In Figure~\ref{fig:q_1_ecc_0.1_SVD_basis}, we show the first four SVD bases.

\begin{figure*}
\includegraphics[width=\textwidth]{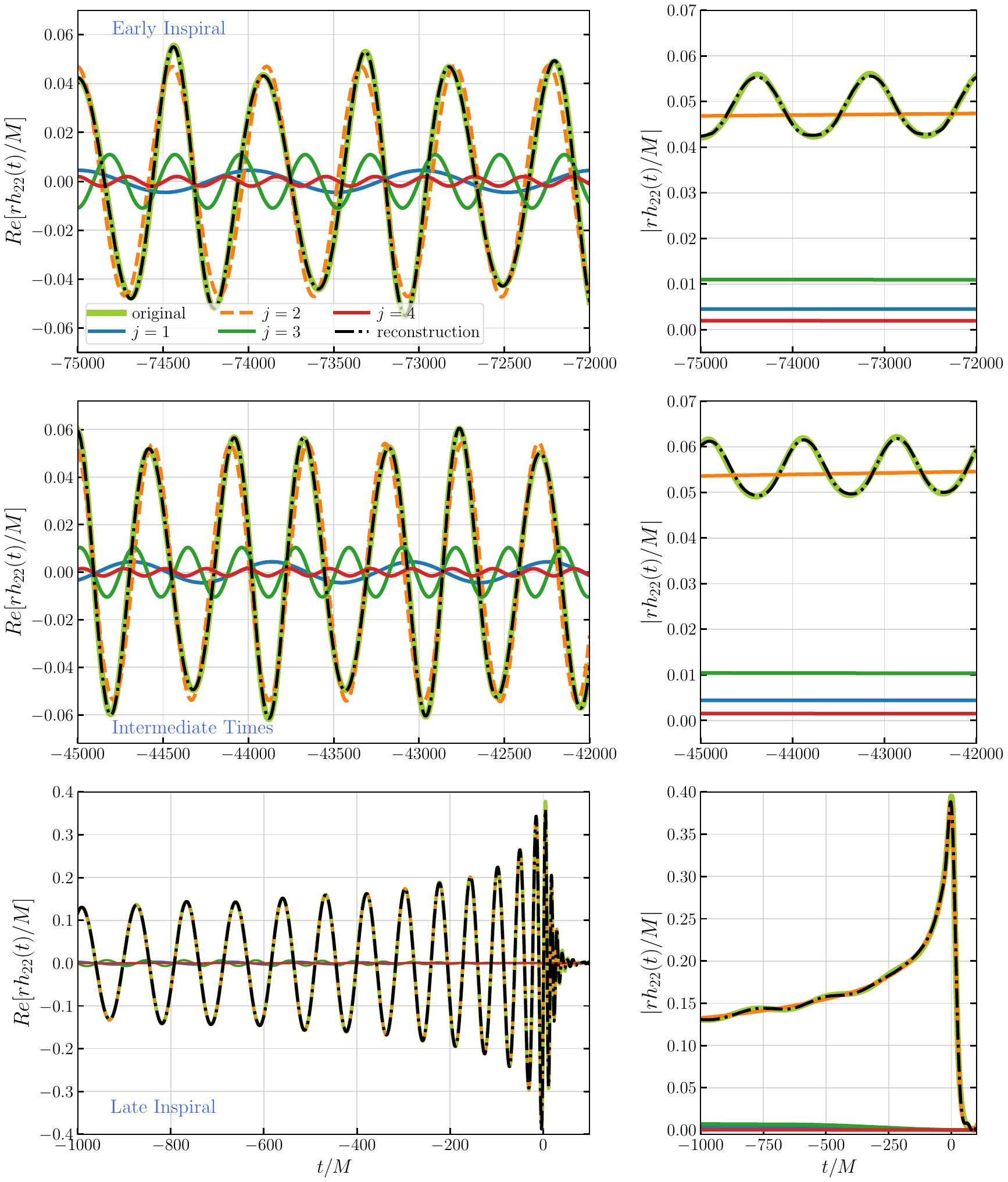}
\caption{We show four eccentric harmonics (and their amplitudes) obtained from the binary with $q=1$, $e_{\rm ref}=0.1$ and $l_{\rm ref}=\pi$ (shown in Fig.~\ref{fig:spectogram} at three different time windows ($[-75000,-72000]M$, $[-45000,-42000]M$, $[-1000,100]M$) for visual inspection. Additionally, we show the original \texttt{TEOBResumS} waveforms (solid green line) along with the reconstructed waveform (dashed black line). More details are in Section~\ref{sec:ecc_harm_example}. 
}
\label{fig:q1_ecc_0.1}
\end{figure*}

We find that different harmonic basis have varying numbers of cycles, indicating different frequency evolutions. Furthermore, all harmonics except the leading one exhibit an overall decreasing amplitude (this is also expected from eccentric harmonics as eccentricity decreases with time). However, we notice that around the merger, all subdominant harmonic bases display a merger-ringdown-like structure. This is unexpected if these harmonics (and therefore basis vectors) are induced purely by eccentricity, as eccentricity is supposed to be negligible near merger for the cases we consider.

The only explanation is that some power from the leading harmonic is spilling into the higher harmonic bases, and we must filter those out. It is important to note that this spill does not affect the dominant $j=2$ harmonic basis, as its amplitude is orders of magnitude larger than the subdominant ones. For most cases we consider, we find that due to numerical noise (originating from the imperfection of the SVD method), we cannot reliably identify more than four or five clean eccentric harmonic bases from the data. We find that a total of 4 eccentric harmonics are good enough for us for eccentricities up to $0.3$. In future sections, we will show that the reconstruction errors with this four harmonics will be very small (Figure~\ref{fig:SVD_recon}). Similar results are obtained in Ref.~\cite{Patterson:2024vbo}.

%==========================================================================
\subsubsection{Smoothing eccentric harmonic basis using PN intuition}
\label{sec:svd_smooth}
To smoothen the eccentric harmonic basis vectors, we perform a simple smoothing exercise. For the phase, we use a spline (from \texttt{scipy.interpolate}~\footnote{\href{https://docs.scipy.org/doc/scipy/reference/interpolate.html}{https://docs.scipy.org/doc/scipy/reference/interpolate.html}}) with a smoothing option to remove some of the extra noise. For the amplitude, we apply the same technique for $j=2$. For the other harmonic bases, we first note that their amplitude is expected to be only a function of eccentricity. Typically, Newtonian and PN calculations suggest that the amplitude of these harmonic bases follows a form~\cite{Yunes:2009yz,VanDenBroeck:2006qu,Arun:2007qv,Seto:2001pg,2012ApJ74537V}:  
\begin{equation}
p_j(e) \approx A e_{\rm t}^n + B e_{\rm t}^{n+2} + C e_{\rm t}^{n+4} + \dots,
\end{equation}
where $e_{\rm t}$ is the eccentricity as function of time. However, we find that we can assume an overall eccentricity dependence of the form:  
\begin{equation}
p_j(e) = a e_{\rm 3PN}^n,
\end{equation}
where $a$ and $n$ are to be determined from the data. As before, we use a 3PN calculation to obtain the eccentricity evolution in time. Since eccentricity tends to zero around the merger, this functional form also ensures that the fitted amplitude naturally decays near the merger. Figure~\ref{fig:q_1_ecc_0.1_SVD_basis} shows the first four SVD bases after smoothing.

%==========================================================================
\subsubsection{Obtaining eccentric harmonics}
\label{sec:ecc_harm_example}
Once we obtain the eccentric harmonic basis, any waveform in the $X$ matrix can be written as:  
\begin{equation}
h_k(t) \approx \sum_{j=1}^{4} h_{k,j}(t),
\end{equation}
where $h_{k,j}(t)$ are the eccentric harmonic modes given by:  
\begin{equation}
h_{k,j}(t) = C_{j}(l_{\rm ref}) e_{j}(t),
\end{equation}
where $C_{j}(l_{\rm ref})$ are the eccentric harmonic coefficients (which depend only on the mean anomaly values, since $q$ and $e_{\rm ref}$ are the same for all), and $e_{j}(t)$ are the normalized eccentric harmonic basis vectors. We obtain the coefficients $C_{j}(l_{\rm ref})$ by projecting each waveform $h_k(t)$ onto the smooth eccentric harmonic basis:
\begin{equation}
C_{j}(l_{\rm ref}) = \langle h_{k,j}(t), e_{j}(t) \rangle = h_{k,j}(t) \cdot e_{j}(t).
\label{eq:cj_coeff}
\end{equation}

For completeness, we present the four eccentric harmonics obtained from the binary with $q=1$ and $e_{\rm ref}=0.1$ (shown in Fig.~\ref{fig:spectogram}) at three different time windows ($[-75000,-72000]M$, $[-45000,-42000]M$, $[-1000,100]M$) in Figure~\ref{fig:q1_ecc_0.1}. 
Notably, at early times, all four harmonics contribute significantly to the waveform, but at late times, the signal is predominantly dominated by the $j=2$ harmonic. 
This occurs because, towards the end of the inspiral, the eccentricity decreases, leading to a suppression of higher-order eccentric harmonics. Additionally, we show the original \texttt{TEOBResumS} waveforms along with the reconstructed waveform using the four harmonics. We demonstrate that this reconstruction effectively recovers the original signal.
We also plot the amplitudes of the individual harmonics on the same figure. Each eccentric harmonic exhibits a monotonic amplitude evolution, yet their superposition—using the four leading harmonics—reproduces the oscillatory amplitude modulation seen in the full 
$(\ell,m)=(2,2)$ mode.

%==========================================================================
\subsubsection{Effect of mean anomaly parameter}
\label{sec:mean_ano_variance}
To find out the effect of the mean anomaly parameter on various eccentric harmonics, we can simply look at the change in coefficient values.
We find that the coefficients exhibit a periodic variation as we change the mean anomaly value. This is probably not surprising given that mean anomaly itself is a periodic parameter. Furthermore, Newtonian calculation hint a sinusoidal-like behavior for these coefficients. Note that each eccentric harmonic basis is a complex-valued time series. Therefore, a negative value in any of the relative coefficients does not mean a diminished contribution.

\begin{figure}
\includegraphics[width=\columnwidth]{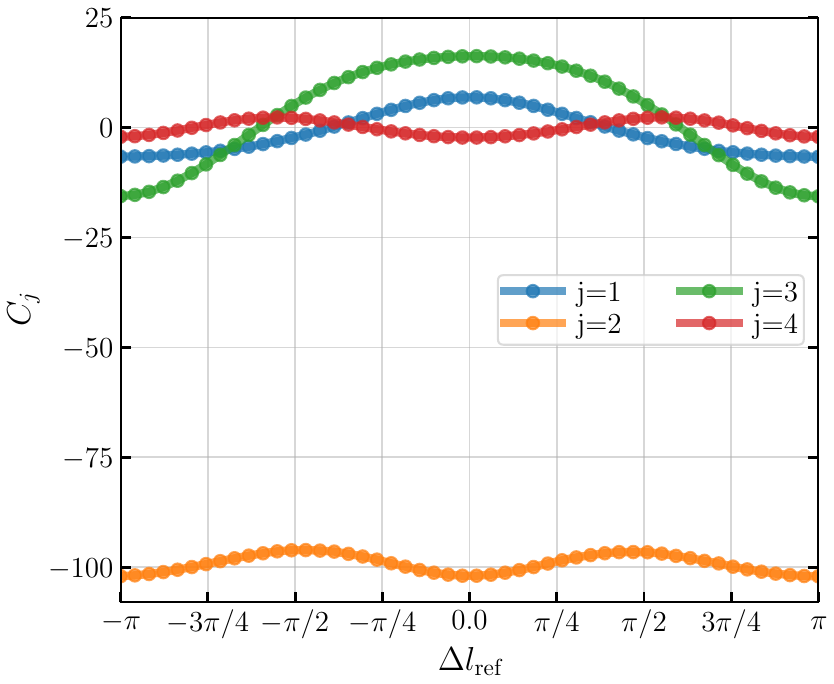}
\caption{We show the variation of the coefficients $C_{j}(l_{\rm ref})$ to the eccentric harmonic basis for four most important eccentric harmonic basis. Details are provided in Section~\ref{sec:mean_ano_variance}.}
\label{fig:mean_ano_variance}
\end{figure}

\begin{figure}
\includegraphics[width=\columnwidth]{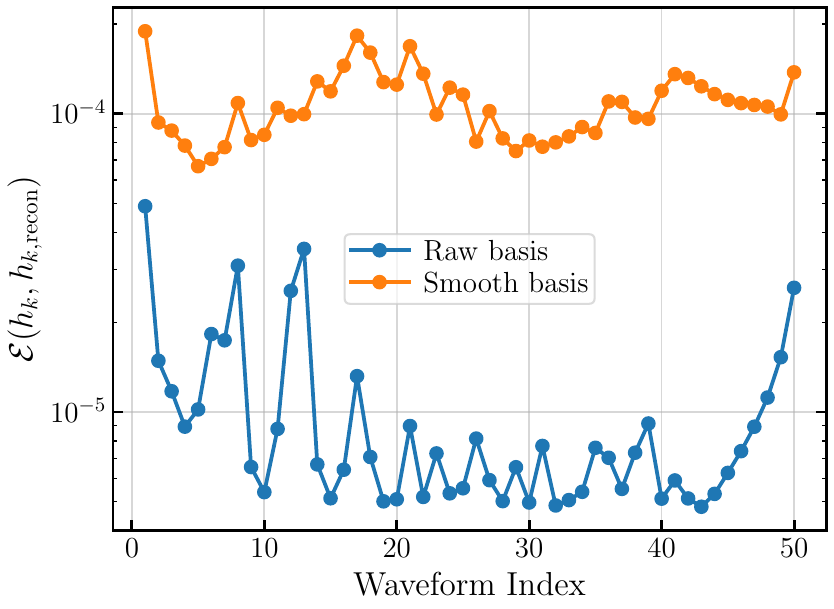}
\caption{We show the reconstruction errors for all 50 waveforms used in the SVD-based method to find the eccentric harmonics. We use four eccentric harmonic bases (both the raw and smooth ones) indexed by $j=[1,2,3,4]$. Details are provided in Section~\ref{sec:reconstruction}.}
\label{fig:SVD_recon}
\end{figure}

%==========================================================================
\subsubsection{Reconstruction accuracy of eccentric harmonic decomposition}
\label{sec:reconstruction}
Now, it is time to ask how much of the original signal can be recovered by the eccentric harmonic decomposition obtained through SVD (and subsequent smoothing). One worry is always whether we would lose a significant part of the signal because of the de-noising. 

To answer these questions, we reconstruct the full signal by summing up the eccentric harmonics obtained by using the original noisy eccentric harmonic basis as well as by using the smooth eccentric harmonic basis vectors. We show the reconstruction errors (Fig.~\ref{fig:SVD_recon}) for all 50 waveforms used to find the eccentric harmonics for $q=1$ and $e_{\rm ref}=0.1$. We find that while the original basis vectors can reconstruct the full waveform with an average $L_2$-norm error of $10^{-5}$, the smooth basis vectors do the same with an average $L_2$-norm error of $10^{-4}$. For most practical purposes such as waveform modeling based on these harmonics or for building a template bank, these errors are considered to be very small.

\begin{figure}
\includegraphics[width=\columnwidth]{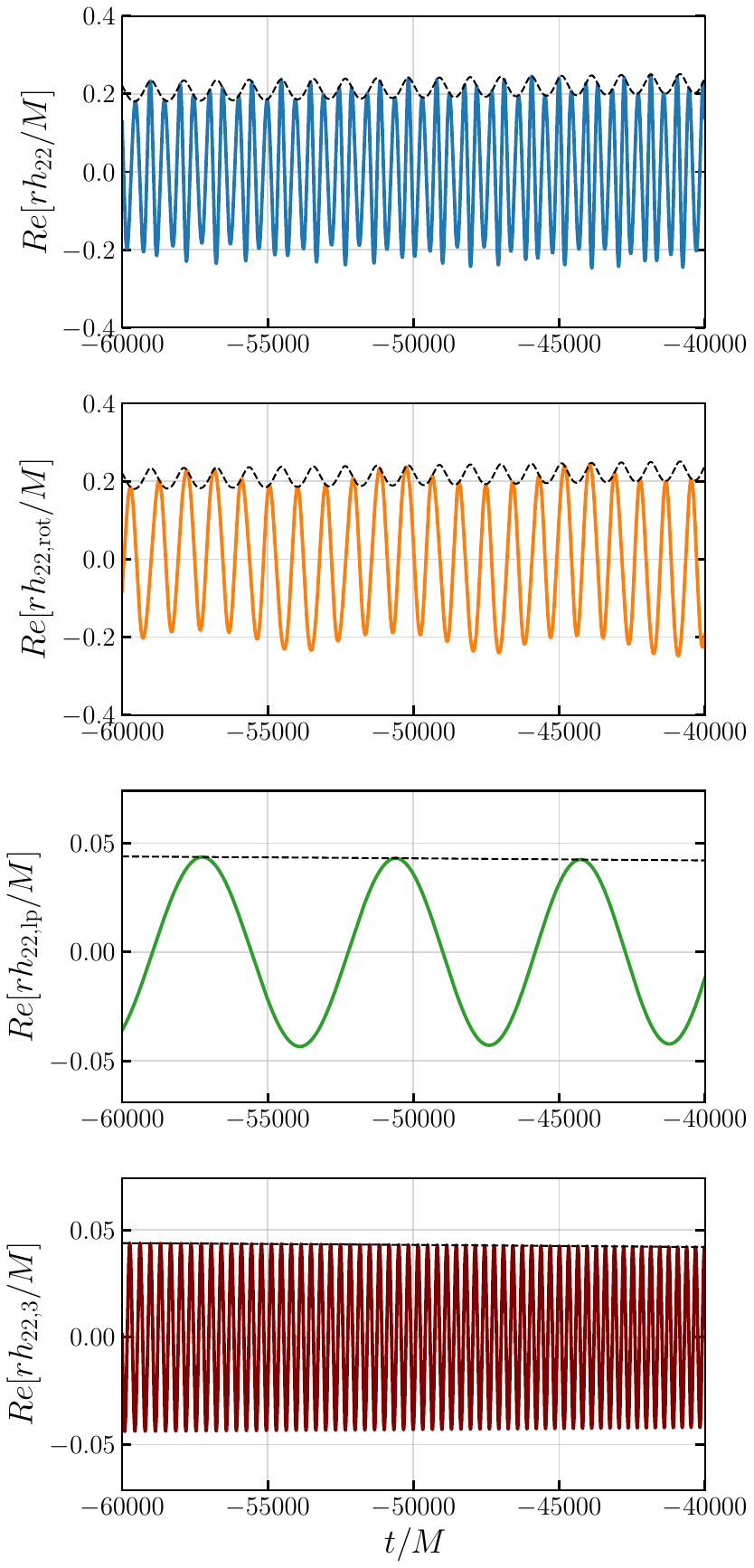}
\caption{Apart from the SVD-based method, we probe another method based on using heterodyning and low-pass filter to extract the harmonics. We demonstrate the steps to extract the $j=3$ eccentric harmonic from the signal shown in Fig.~\ref{fig:spectogram} using the filter-based method. For visual clarity, we zoom into a smaller time window of $[-60000,-40000]M$. The first panel shows the original signal, the last panel presents the extracted $j=3$ eccentric harmonic. Middle two panels show the modified waveforms in the intermediate steps just before and after applying the low-pass filter. We follow similar steps to extract any other eccentric harmonic. For each panel, we also show the amplitude of the waveform as a dashed black line. Note that while the original signal has oscillatory amplitude due to the superposition of multiple eccentric harmonics, amplitude of the extracted $j=3$ eccentric harmonic is monotonic indicating the signal has only one eccentric harmonic. More details are in Section~\ref{sec:filter_method}.}
\label{fig:filter_example}
\end{figure}

%==========================================================================
\subsubsection{Limitation of the SVD-based method}
One potential limitation of the SVD-based method is the computational cost. For example, to find out the $\omega_{\rm orb, left}$, we need to evaluate \texttt{TEOBResumS} waveform around 30 times. Then we need to generate about 50 waveforms at equally spaced (ideally) mean anomaly values to make sure the inputs to SVD have a good representation of the full range of mean anomaly values. All these become computationally expensive, especially when we try to generate really long waveforms (for example, $75000M$ here) to make sure it starts from $20$ Hz in the LIGO sensitivity band. One potential solution could be to build a reduced order surrogate model for the eccentric harmonics in the future.

\begin{figure*}
\includegraphics[width=\textwidth]{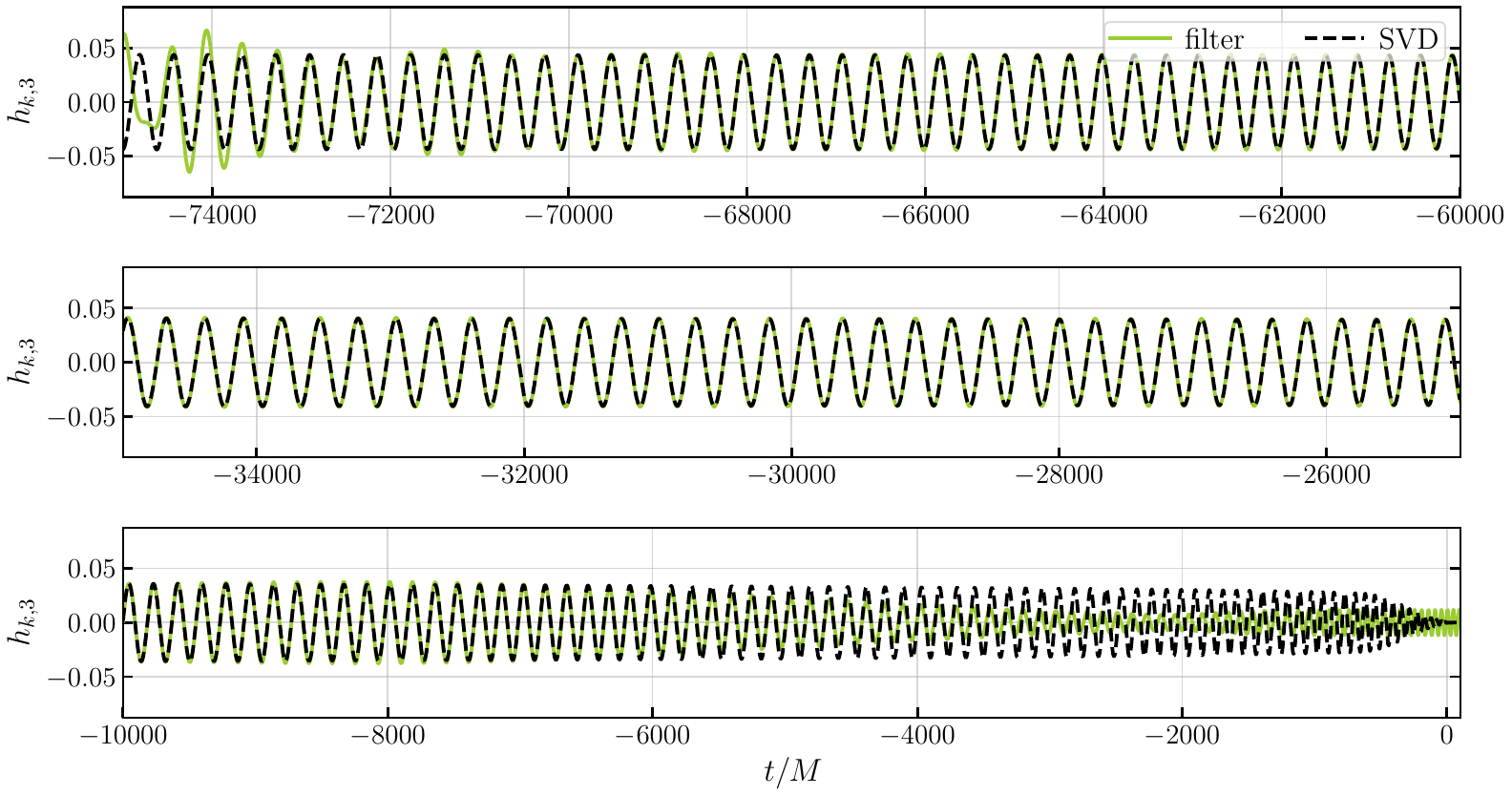}
\caption{We show the comparison between the $j=3$ eccentric harmonic extracted using the filter-based method (green solid line) and the SVD-based method (black dashed line) at three different time windows: early inspiral ($[-75000, -60000]M$), intermediate time ($[-35000, -25000]M$), and late inspiral-merger-ringdown ($[-10000, 100]M$). Details are provided in Section~\ref{sec:filter_vs_svd}.}
\label{fig:filter_vs_svd}
\end{figure*}

%==========================================================================
%==========================================================================
\subsection{Alternative method to extract eccentric harmonics using low-pass filters}
\label{sec:filter_method}
We now develop a complementary framework based on signal processing that exploits the approximate phase (and frequency) relations between eccentric harmonics, as expected in Newtonian scenarios, and extracts individual eccentric harmonics from the full signal. The main idea is to use heterodyning followed by a low-pass filter to extract eccentric harmonics one at a time. This framework, while having limitations (which we will discuss in the following sections), will serve as a sanity check for the method proposed in Section~\ref{sec:svd_method} that relies on the SVD. 

First, we use the quadrupolar mode of the radiation to obtain a monotonic average instantaneous orbital frequency $\omega_{\rm orb, \rm avg}$. Details are in Appendix~\ref{sec:avg_orb_phase}.

To extract a specific eccentric harmonic characterized by $j$, we first apply a phase rotation to the full waveform (which technically contains all eccentric harmonics) and remove the leading approximate expected phase of the $j$-th eccentric harmonic:
\begin{align}
h_{22, \rm rot}(t) = h_{22}(t) e^{-ij \phi_{\rm orb, \rm avg}(t)}.
\label{eq:ecc_phase_rot}
\end{align}
This phase rotation ensures that the modified waveform mode now contains frequencies that are mostly unrelated to the harmonic of interest.

Next, we construct a low-pass filter that eliminates any frequency content above a certain frequency threshold in an input signal. We use the \texttt{scipy.signal} module~\footnote{\href{https://docs.scipy.org/doc/scipy/reference/signal.html}{https://docs.scipy.org/doc/scipy/reference/signal.html}} for our purposes. The maximum allowed frequency is set to be slightly smaller than the average initial instantaneous orbital frequency. This value also matches the expected approximate difference in frequencies between two eccentric harmonics. We then convolve the low-pass filter with the modified waveform mode. This gives us a signal that contains only the $j$-th eccentric harmonic, but with its phase rotated.
\begin{equation}
    h_{22, \rm lp}(t) = (h_{22, \rm rot} * g)(t) = \int_{-\infty}^{\infty} h_{22, \rm rot}(\tau) g(t - \tau) \, d\tau
\end{equation}
where $*$ denotes convolution and $g(t)$ is the impulse response of the low-pass filter. The impulse response $g(t)$ is the inverse Fourier transform of the frequency response $G(f)$. Details of the low-pass filter is provided in Appendix~\ref{sec:lpfilter}.

Finally, we modify the filter output by undoing the initial phase rotation. This provides us with the $j$-th eccentric harmonic.
\begin{align}
h_{22,j}(t) = h_{22,\rm lp}(t) e^{ij \phi_{\rm orb, \rm avg}(t)}.
\end{align}

We demonstrate the steps in the filter-based method in Fig.~\ref{fig:filter_example} for the signal shown in Fig.~\ref{fig:spectogram}. In particular, for this demonstration, we focus on the $j=3$ eccentric harmonic and zoom into a smaller time window of $[-60000,-40000]M$ to inspect the waveform cycles properly. The original quadrupolar waveform (which contains all the eccentric harmonics) is shown in the first panel. Notably, the amplitude (black dashed line) exhibits oscillatory behavior as the full amplitude originates from the superposition of many eccentric harmonics. After applying a phase rotation, according to Eq.~(\ref{eq:ecc_phase_rot}), to this signal to remove the leading phase of the $j=3$ eccentric harmonic, we obtain $h_{22,\rm rot}$ --- which has fewer cycles than before. However, it still retains an oscillatory amplitude, as all eccentric harmonics remain present in the signal. Next, we pass $h_{22,\rm rot}$ through a low-pass filter (whose frequency response is depicted in Fig.~\ref{fig:filter_example}), which eliminates all eccentric harmonics except for $j=3$. As a result, the amplitude of the filtered signal $h_{22,\rm lp}$ is significantly smaller compared to the original waveform in the first two panels. Finally, we restore the originally removed leading phase of the $j=3$ harmonic, yielding $h_{22,j=3}$ (shown in the lower panel). This waveform no longer exhibits oscillatory amplitude, indicating that it contains only the single $j=3$ eccentric harmonic. A similar procedure is followed to obtain the eccentric harmonics associated with other $j$.

\begin{figure*}
\includegraphics[width=\textwidth]{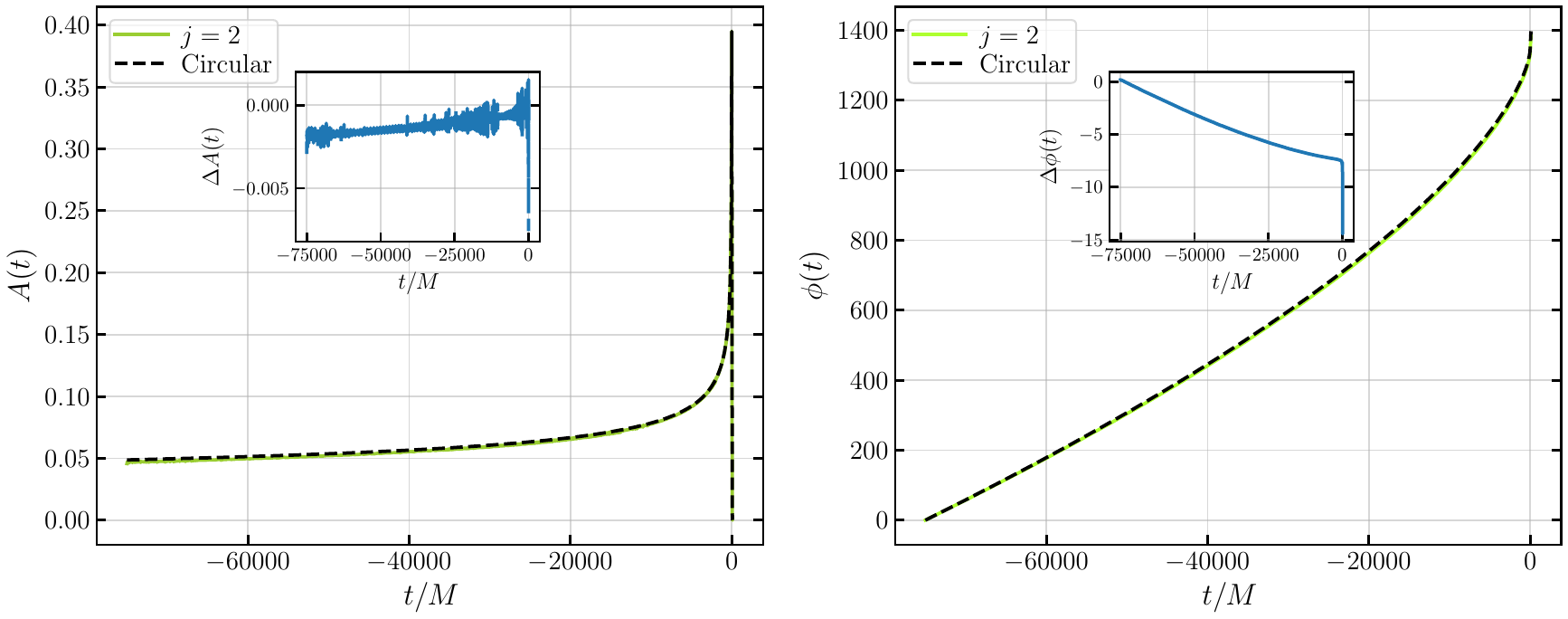}
\caption{We show the comparison between the amplitude (left panel) and phase (right panel) of the $j=2$ eccentric harmonic for $q=1$ and $e_{\rm ref}=0.1$ with the circular expectations. The differences are shown in the inset. Details are provided in Section~\ref{sec:j2_vs_circular}.}
\label{fig:j2_vs_cir}
\end{figure*}

While the approach using the filter provides an easier and faster way to extract the eccentric harmonics, it is susceptible to the Gibbs phenomenon due to the sharp discontinuities at both ends of the waveform time window. Due to this, we can only extract clean harmonics at the intermediate times. Details of this limitation is provided in Appendix~\ref{sec:filter_limitation}.

%==========================================================================
%==========================================================================
\subsection{Comparison between SVD- and filter-based methods}
\label{sec:filter_vs_svd}

We conclude this section by providing a one-to-one comparison between the eccentric harmonics obtained through the SVD- and filter-based methods. For the SVD-based method, we have used 50 waveforms with different mean anomaly values to extract the harmonic basis. For the comparison, we however choose the parameters corresponding to the reference waveform, i.e., we stick to $l_{\rm ref}=\pi$. As discussed previously, the filter-based method suffers from the Gibbs phenomenon in the early and late times. We will therefore perform the comparison only in the intermediate time window where both methods provide clean harmonics. For the waveform shown in Fig.~\ref{fig:spectogram}, this window is $[-65000M, -15000M]$ for all eccentric harmonics.

In Fig.~\ref{fig:filter_vs_svd} we show the eccentric harmonics obtained from these two methods for $j=3$ for demonstration purposes. Note that these two methods yield harmonics that visibly match each other in the intermediate times. It therefore acts as a confirmation for both the methods that they are extracting the same physics. Furthermore, at this intermediate time range, we calculate the time-domain error between $j=3$ harmonics obtained from these two methods. We find this error to be $0.00063$, which is very small. We observe a similar degree of match for other eccentric harmonic modes as well.

Another noteworthy aspect is the differences in computational cost between these two methods. The filter-based method is faster, taking only $5$ to $10$ seconds for each case, while the SVD-based method takes around $300$ to $600$ seconds depending on the parameter values.

%==========================================================================
%==========================================================================
%==========================================================================
\section{Phenomenology of eccentric harmonics}
\label{sec:phenomenology}
Now that we have presented two different ways to extract the eccentric harmonics for any non-spinning eccentric waveforms and demonstrated that these two methods provide the same harmonics, we now shift our focus to the phenomenology of these harmonics. In particular, we try to understand whether they follow any simple structure that may prove helpful in future modeling. We therefore focus on the amplitudes, phase, and instantaneous orbital frequencies for these modes. We restrict our studies to $j\in\{1,2,3,4\}$ as they are the most important ones and can be obtained with fewer errors from the data. Furthermore, we focus only on binaries with $q=1$ and $e_{\rm ref}=0.1$, as in most of the paper. Changing the binary parameters does not alter the phenomenology we study.

%==========================================================================
%==========================================================================
\subsection{Comparison between $j=2$ eccentric harmonic and circular waveform}
\label{sec:j2_vs_circular}
First, we inspect the $j=2$ eccentric harmonic. Note that this harmonic is expected to be closest to the circular case. To check how the $j=2$ eccentric harmonic differs from the circular waveform, we generated the circular expectation (i.e. with $e_{\rm ref}=0.0$) for $q=1$ with an initial orbital frequency such that the length of the waveform is $75000M$. 

We then compare their amplitudes and phases (Fig.~\ref{fig:j2_vs_cir}). We find that overall, the circular waveform matches closely with the $j=2$ eccentric harmonic. However, due to eccentricity, the $j=2$ harmonic has additional non-negligible phases related to perihelion precession. Furthermore, its amplitude also exhibits a slight difference. We calculate their time-domain error for the full waveform and find it to be quite large (0.66, to be exact). This means that it is not possible to approximate the $j=2$ harmonic by a circular waveform with the same mass ratio and spin values.

%==========================================================================
%==========================================================================
\subsection{Hierarchy of eccentric harmonics}
\label{sec:ecc_harm_hierarchy}
\begin{figure}
\includegraphics[width=\columnwidth]{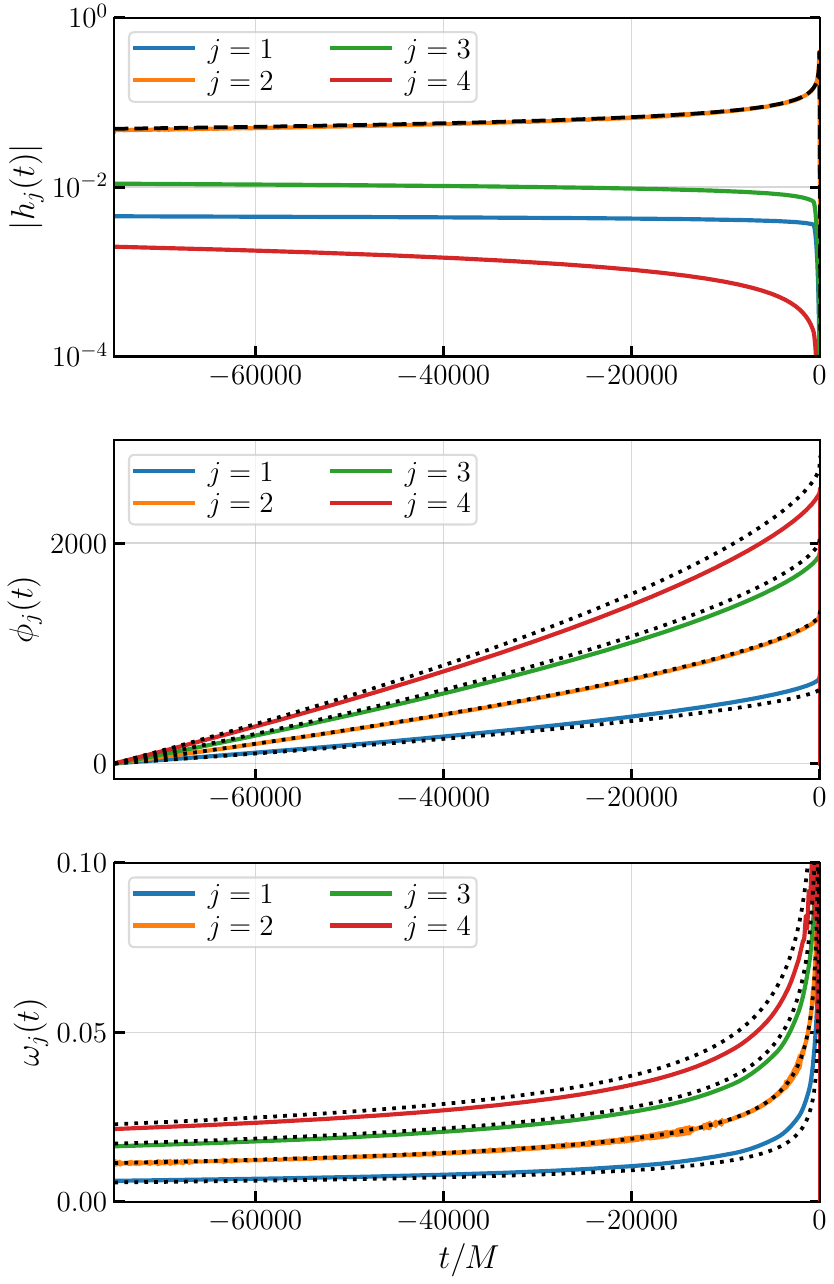}
\caption{We show the amplitudes (upper panel), phases (middle panel), and instantaneous frequencies (lower panel) for eccentric harmonics with $j = [1,2,3,4]$. For the amplitude, we also plot the amplitude of the circular waveform (as a dashed line). For the phases and frequencies, we plot the naive $j$-scaled expectations (as shown in Eq.~(\ref{eq:phase_j_relation}) and Eq.~(\ref{eq:freq_j_relation})) as dotted lines. Details are provided in Section~\ref{sec:ecc_harm_hierarchy}.}
\label{fig:ecc_harm_hierarchy}
\end{figure}

Next, we plot the amplitudes (upper panel), phases (middle panel), and instantaneous frequencies (lower panel) for all four eccentric harmonics in Fig.~\ref{fig:ecc_harm_hierarchy}. For the amplitude, we also plot the amplitude of the circular waveform (as a dotted line). For the phases and frequencies, we plot the naive $j$-scaled expectations (as shown in Eq.~(\ref{eq:phase_j_relation}) and Eq.~(\ref{eq:freq_j_relation})) as dotted lines. We reach three interesting conclusions:
\begin{itemize}
    \item At all times, the $j=2$ amplitude remains the loudest, followed by $j=3$, $j=1$, and $j=4$. Other eccentric harmonics neglected in this study will have even smaller amplitudes. Note that the difference between the dominant $j=2$ and $j=4$ amplitudes is already two orders of magnitude.
    \item Eccentric harmonic phases cannot simply be assumed to be $j \phi_{\rm orb}$ or $j \phi_{\rm orb, cir}$. Due to eccentricity, there are additional phase corrections that these eccentric harmonics inherit.
    \item For the same reason, simple frequency expectations such as $j \omega_{\rm orb}$ are not valid for eccentric harmonics in relativistic scenarios. There are corrections that need to be taken into account.
\end{itemize}

\begin{figure}
\includegraphics[width=\columnwidth]{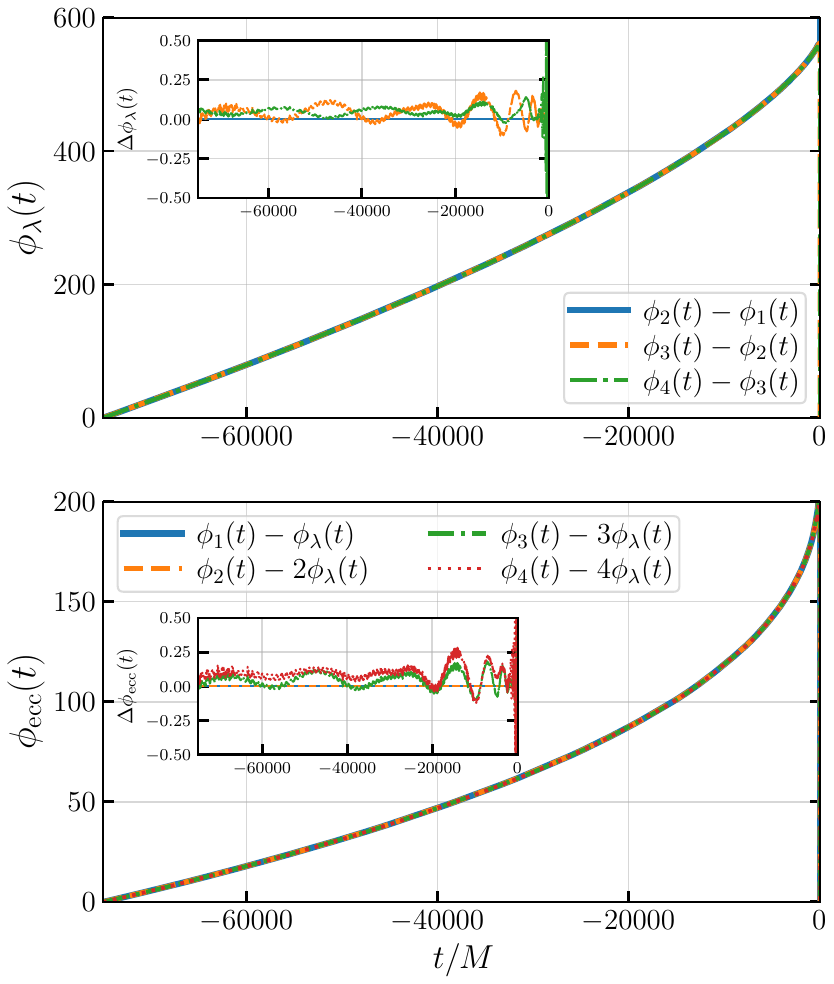}
\caption{We show the phase relation between different eccentric harmonics for the signal shown in Fig.~\ref{fig:spectogram}. The upper panel shows $\phi_{\lambda}(t)$ while the lower panel presents $\phi_{\rm ecc}(t)$ computed from different extracted modes. We show the residuals in the inset. Details are provided in Section~\ref{sec:ecc_harm_hierarchy}. 
}
\label{fig:ecc_harm_phase_relation}
\end{figure}

We figure out that the eccentric harmonics follow a phase and frequency structure that consists of two different components. The first component is related to the orbital phase, while the second component is solely influenced by eccentricity. We can formally express these relations as:
\begin{align}
\phi_{\ell m,j}(t) &\approx j \phi_{\lambda}(t) + \phi_{\rm ecc}(t),\\
\omega_{\ell m,j}(t) &\approx j \omega_{\lambda}(t) + \omega_{\rm ecc}(t),\\
f_{\ell m,j}(t) &\approx j f_{\lambda}(t) + f_{\rm ecc}(t),
\end{align}
where the phase $\phi_{\lambda}(t)$ (and its related instantaneous frequency $\omega_{\lambda}(t)$ and GW frequency $f_{\lambda}(t)$) corresponds to the secular part of the orbital phase in the circular case, scaling with the value of $j$. However, the second component is only affected by eccentricity and remains independent of $j$. We refer to this as the eccentric phase (or eccentric instantaneous frequency/eccentric GW frequency). For the $j=3$ mode, we also observe an additional $\pi$ offset in the phase.

In Fig.~\ref{fig:ecc_harm_phase_relation}, we demonstrate the proposed relation in action. First, we show that if we compute the phase difference between the phase of any two subsequent eccentric harmonics, they all match as expected according to the above relation. This difference gives the secular phase $\phi_{\lambda}(t)$, which scales with the value of $j$. Then, within each eccentric harmonic, whatever phase remains is the same for all $j$—this is the eccentric phase $\phi_{\rm ecc}(t)$. We illustrate this in the lower panel of the same figure. Note that while generating this plot, we account for the initial $\pi$ phase offset in the $j=3$ mode.

%==========================================================================
%==========================================================================
\subsection{Comparison between $\phi_{\lambda}(t)$, $\phi_{\rm ecc}(t)$ and 3PN $l_{3PN}(t)$}
\label{sec:phase_comparison}
We now aim to connect the extracted phase parameters $\phi_{\lambda}(t)$ and $\phi_{\rm ecc}(t)$ 
to other phase parameters of interest. For example, the most commonly available phase parameter for eccentric waveforms is the average orbital phase of the binary, $\phi_{\rm orb, avg}$, as presented in Section~\ref{sec:avg_orb_phase}. Additionally, using post-Newtonian (PN) calculations, we can obtain the expected mean anomaly phase of the binary, $l_{\rm 3PN}(t)$, as presented in Ref.~\cite{Moore:2016qxz}. 

We find that $\phi_{\rm orb, avg}$ is significantly larger than both $\phi_{\lambda}(t)$ and $\phi_{\rm ecc}(t)$. Interestingly, $\phi_{\lambda}(t)$ closely follows the evolution of $l_{\rm 3PN}(t)$, suggesting that these two quantities are essentially the same. In other words:
\begin{equation}
    l(t) \approx \phi_{\lambda}(t).
\end{equation}
Any observed differences are likely due to the absence of higher-order PN terms. Additionally, our 3PN expression for $l_{\rm 3PN}(t)$ is primarily valid for low eccentricities ($e \lesssim 0.2$). Nonetheless, this result provides a new measure of the mean anomaly parameter for binary systems.

Furthermore, we also calculate the expected 3PN values of the periastron advance $k_{\rm 3PN}(t)$ using PN expressions. To calculate that, we use the average orbital frequencies obtained in Section~\ref{sec:avg_orb_phase}. We find that $k_{\rm 3PN}(t)$ is too small to account for the $\phi_{\rm ecc}(t)$ component. While $k_{\rm 3PN}(t)l_{\rm 3PN}(t)$ values are closer to $\phi_{\rm ecc}(t)$ their shape does not match well. 

\begin{figure}
\includegraphics[width=\columnwidth]{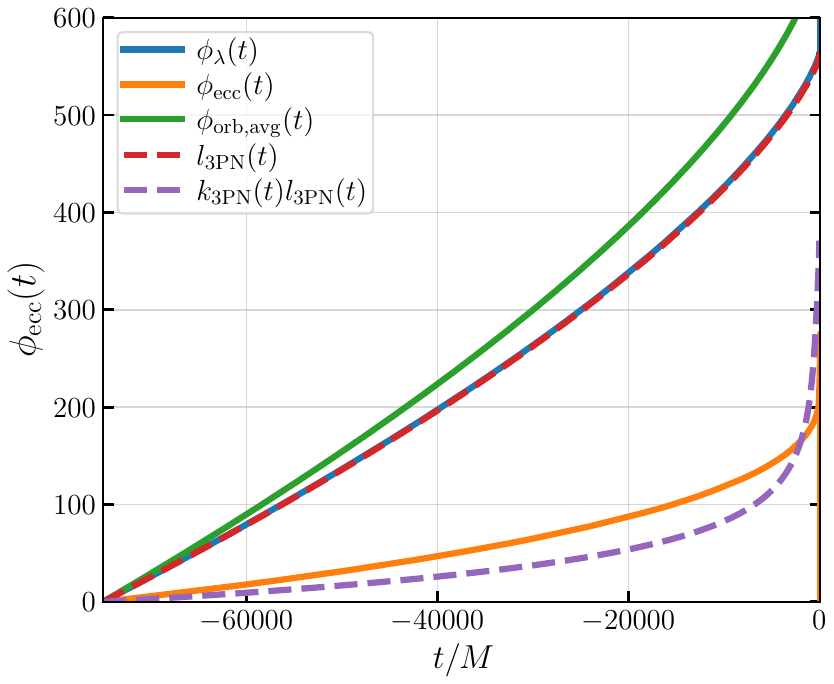}
\caption{We present the extracted phases $\phi_{\lambda}(t)$ and $\phi_{\rm ecc}(t)$ for the binary shown in Fig.~\ref{fig:spectogram}. For comparison, we also display the average orbital phase $\phi_{\rm orb, avg}$, as discussed in Section~\ref{sec:avg_orb_phase}. Additionally, we plot the expected mean anomaly phase $l_{\rm 3PN}(t)$ and periastron advance $k_{\rm 3PN}(t)$ using a 3PN calculation. Details are provided in Section~\ref{sec:phase_comparison}.}
\label{fig:delta_mean_ano_variance}
\end{figure}

\begin{figure*}
\includegraphics[width=\textwidth]{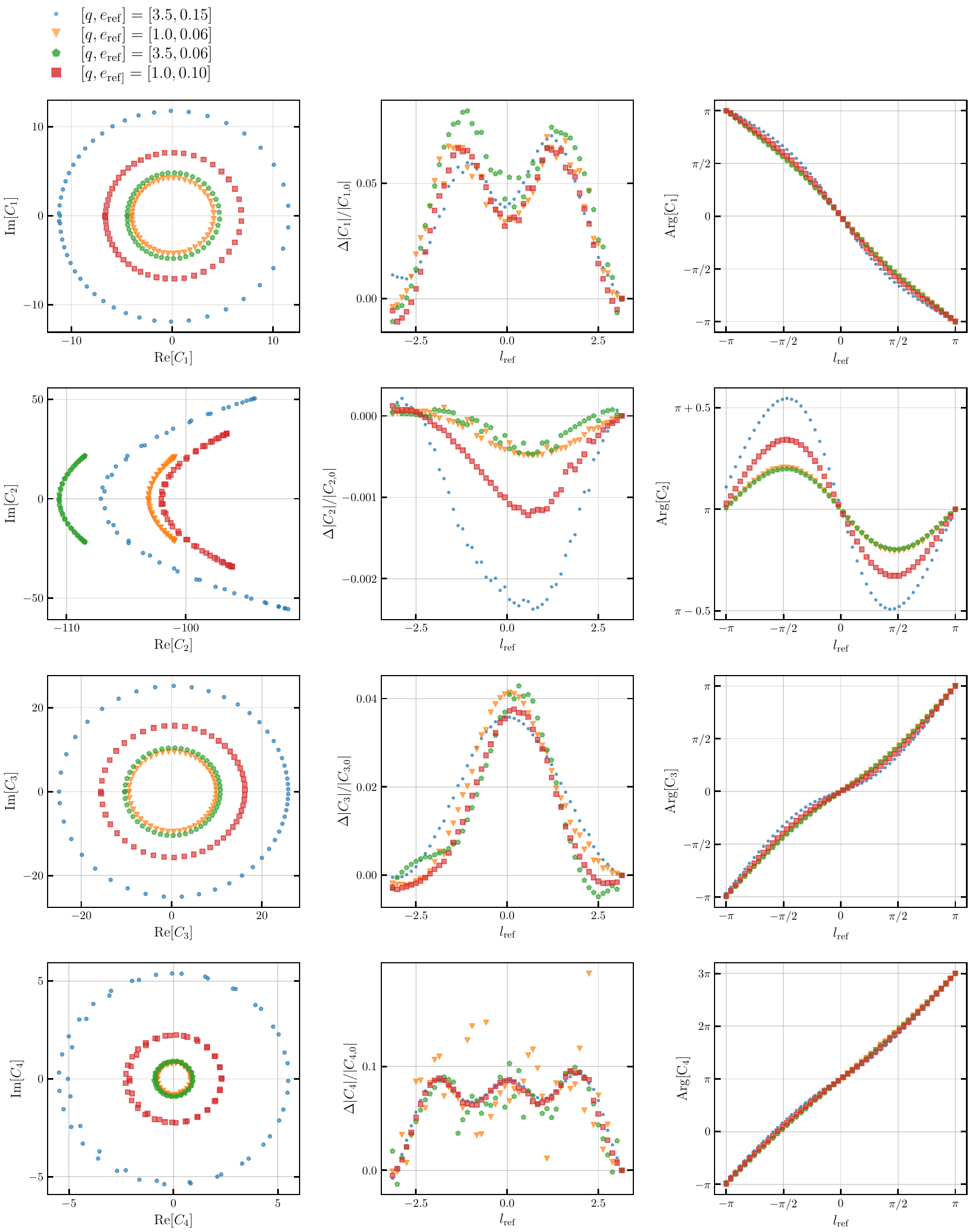}
\caption{We present the SVD coefficients $C_{j}(l_{\rm ref})$ for four different systems, $[q,e_{\rm ref}] = \{[3.5,0.15], [1.0,0.06], [3.5,0.06], [1.0,0.1]\}$, representing different regions of the parameter space. The left panels display the real and imaginary parts of the coefficients, the middle panels show the relative amplitudes, and the right panels present the arguments of the coefficients. Details are provided in Section~\ref{sec:model_mean_anomaly}.}
\label{fig:relative_Cj_vs_lref_combined}
\end{figure*}

%==========================================================================
%==========================================================================
\section{Modeling the effect of mean anomaly}
\label{sec:model_mean_anomaly}
Finally, we shift our focus to modeling the mean anomaly direction, aiming to identify any simple relationships or universal features that could simplify the modeling process. To investigate this, we examine the SVD coefficients $C_{j}(l_{\rm ref})$ (as defined in Eq.~(\ref{eq:cj_coeff})). Note that, these coefficients are complex numbers and therefore have both amplitudes and phases. We, therefore, need to understand the phenomenology of both amplitudes and phases separately and model them. 

\subsection{Modeling the amplitudes of the harmonic coefficients}
\label{sec:model_svd_amp}
Figure~\ref{fig:relative_Cj_vs_lref_combined} presents these coefficients for four representative systems, $[q,e_{\rm ref}] = \{[3.5,0.15], [1.0,0.06], [3.5,0.06], [1.0,0.1]\}$, covering different regions of the parameter space. Specifically, the left panels show the real and imaginary parts of the coefficients, the middle panels display their relative amplitudes with respect to those at $l_{\rm ref}=\pi$, and the right panels illustrate the argument (phase) of the SVD coefficients for $j = [1,2,3,4]$. We define the relative amplitudes as:
\begin{equation}
\Delta |C_{j}|/|C_{j,0}| = \frac{|C_{j}(l_{\rm ref})| - |C_{j}(l_{\rm ref}=\pi)|}{|C_{j}(l_{\rm ref}=\pi)|}.
\end{equation}

We observe that for $j=2$, the amplitude of the SVD coefficients remains nearly unchanged with varying mean anomaly, with a maximum relative deviation of only $0.2\%$ (with respect to the value at $l_{\rm ref}=\pi$) across the cases shown in Figure~\ref{fig:relative_Cj_vs_lref_combined}. \textit{This suggests that $C_{2}(l_{\rm ref})$ can be approximated as constant for different mean anomaly values within the same $q$ and $e_{\rm ref}$:}
\begin{equation}
\Delta |C_{2}|/|C_{2,0}|(l_{\rm ref}) = 0.
\end{equation}

For the other three harmonics, we find an interesting trend. While the amplitudes of the coefficients exhibit noticeable variations with mean anomaly (cf. Figure~\ref{fig:mean_ano_variance}), their relative changes with respect to $l_{\rm ref}=\pi$ appear to follow a similar pattern across all four systems. We verify this trend for other values of $q$ and $e_{\rm ref}$, leading us to decide to model this universal relative change in the SVD coefficient amplitudes as a function of mean anomaly.

\begin{figure}
\includegraphics[width=\columnwidth]{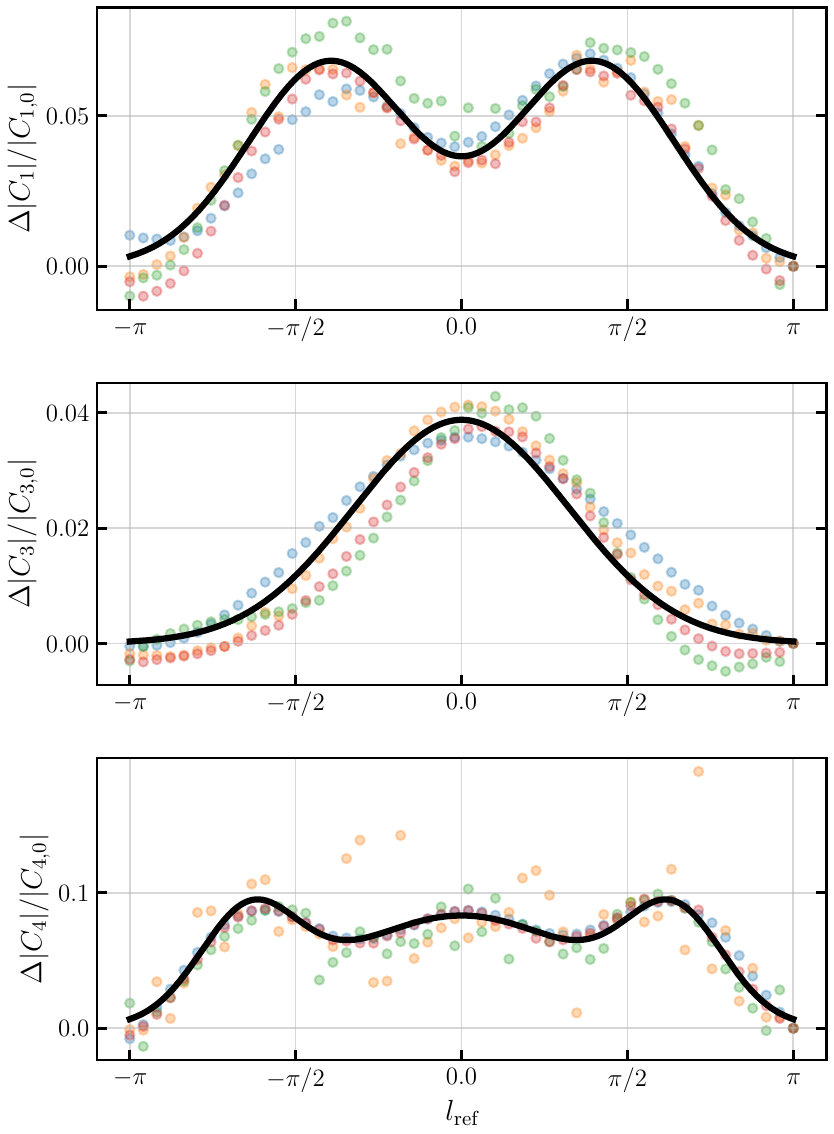}
\caption{We present the relative amplitudes $\Delta |C_{j}|/|C_{j,0}|$ for $j=[1,3,4]$ as differently colored circles for the four systems considered in Figure~\ref{fig:relative_Cj_vs_lref_combined}, along with the corresponding fits shown as solid black lines. Details are provided in Section~\ref{sec:model_mean_anomaly}.}
\label{fig:cj_amplitude_fits}
\end{figure}

At this point, it is important to note that the choice of the reference mean anomaly is arbitrary. We select $l_{\rm ref}=\pi$ as our reference because current EOB models (e.g., \texttt{TEOBResumS}) adopt a similar convention. However, choosing a different reference mean anomaly would yield equivalent results.  

We also note that the relative amplitudes $\Delta |C_{j}|/|C_{j,0}|$ for $j=[1,3,4]$, as shown in Figures~\ref{fig:relative_Cj_vs_lref_combined} and~\ref{fig:cj_amplitude_fits}, exhibit some variance, primarily due to numerical errors in our harmonic extraction procedures. However, we identify the key underlying structures and fit $\Delta |C_{j}|/|C_{j,0}|$ using a double Gaussian for $j=1$, a single Gaussian for $j=3$, and a triple Gaussian function for $j=4$. In all our fits, we enforce symmetry around $l_{\rm ref}=0$. The fitted curves are shown as solid black lines in Figure~\ref{fig:cj_amplitude_fits}.
Best-fit function for $j=1$ is:
\begin{equation}
\Delta |C_{1}|/|C_{1,0}|(l_{\rm ref}) = a_1 \times \left( e^{-\frac{(l_{\rm ref} - x_1)^2}{2\sigma_{1}^2}} + e^{-\frac{(l_{\rm ref} + x_1)^2}{2\sigma_{1}^2}} \right)
\end{equation}
with $a_1=0.07229622$, $x_1=1.27566643$ and $\sigma_1=0.81916535$.
For $j=3$, we get:
\begin{equation}
\Delta |C_{3}|/|C_{3,0}|(l_{\rm ref}) = a_3 \times e^{-\frac{l_{\rm ref}^2}{2\sigma_{3}^2}}
\end{equation}
with $a_3=0.03901312$ and $\sigma_3=1.03484316$.
Finally, for $j=4$, we obtain:
\begin{equation}
\Delta |C_{4}|/|C_{4,0}|(l_{\rm ref}) = A_s \left( e^{-\frac{(x - x_4)^2}{2 \sigma_s^2}} + e^{-\frac{(x + x_4)^2}{2 \sigma_s^2}} \right) 
+ A_c e^{-\frac{x^2}{2 \sigma_c^2}}
\end{equation}
with $x=l_{\rm ref}$, side Gaussian amplitude $A_s = 0.07305414$, central Gaussian amplitude $A_c = 0.09628742$,  side Gaussian peak location $x_4 = 2.06651346$, and Gaussian standard deviations  $\sigma_s = 0.48531689$ and $\sigma_c = 1.41547521$.

\begin{figure*}
\includegraphics[width=\textwidth]{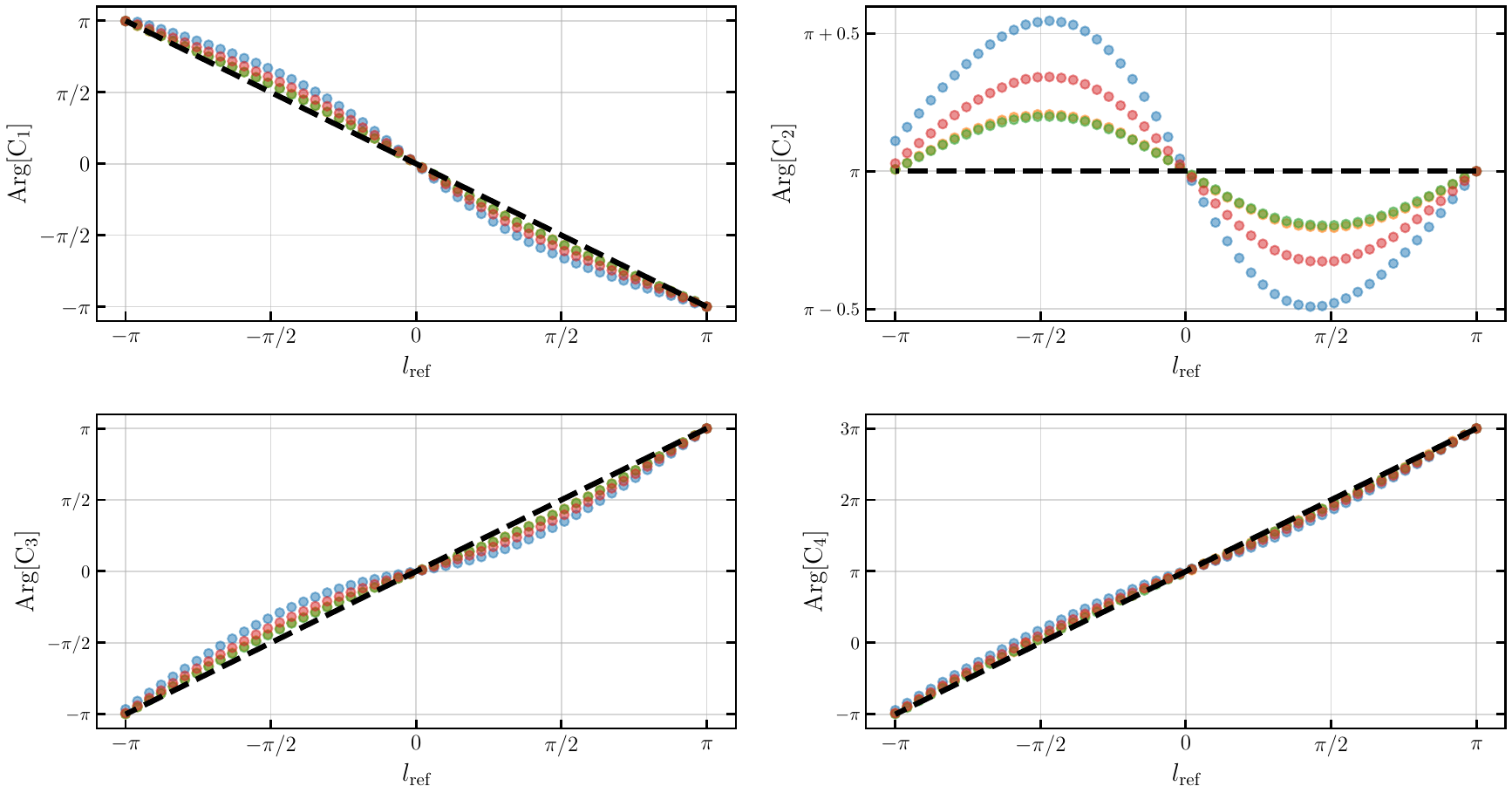}
\caption{We present the arguments of SVD coefficients $C_{j}$ for $j=[1,2,3,4]$ as differently colored circles for the four systems considered in Figure~\ref{fig:relative_Cj_vs_lref_combined}, along with the corresponding leading order fits shown as dashed black lines. Details are provided in Section~\ref{sec:model_mean_anomaly}.}
\label{fig:Cj_arg_linear}
\end{figure*}

\subsection{Modeling the phases of the harmonic coefficients}
\label{sec:model_svd_phase}
Next, we investigate the argument (or phase) part of the SVD coefficients $C_j(l_{\rm ref})$. A quick inspection of Figure~\ref{fig:relative_Cj_vs_lref_combined} suggests that there are two distinct structures in the phases: a leading-order trend and a secondary trend. The leading-order behavior appears to be constant (for $j=2$) or linear (for $j=[1,3,4]$). The full behavior can therefore be written as:
\begin{equation}
{\rm Arg}[C_j](l_{\rm ref}) = \phi_{j,c}^{\rm leading}(l_{\rm ref}) + \phi_{j,c}^{\rm secondary}(l_{\rm ref}),
\end{equation}
where $\phi_{j,c}^{\rm leading}$ represents the dominant phase contribution, and $\phi_{j,c}^{\rm secondary}$ accounts for additional corrections. 
We first model the leading-order behavior and find the following relationships:
\begin{align}
\phi_{1,c}^{\rm leading}(l_{\rm ref}) &= -l_{\rm ref},\\
\phi_{2,c}^{\rm leading}(l_{\rm ref}) &= \pi,\\
\phi_{3,c}^{\rm leading}(l_{\rm ref}) &= l_{\rm ref},\\
\phi_{4,c}^{\rm leading}(l_{\rm ref}) &= 2l_{\rm ref} + \pi.
\end{align}
Figure~\ref{fig:Cj_arg_linear} shows the arguments of $C_{j}$ for $j=[1,2,3,4]$ as differently colored circles for the four systems considered in Figure~\ref{fig:relative_Cj_vs_lref_combined}, with the corresponding leading-order fits shown as dashed black lines.

\begin{figure}
\includegraphics[width=\columnwidth]{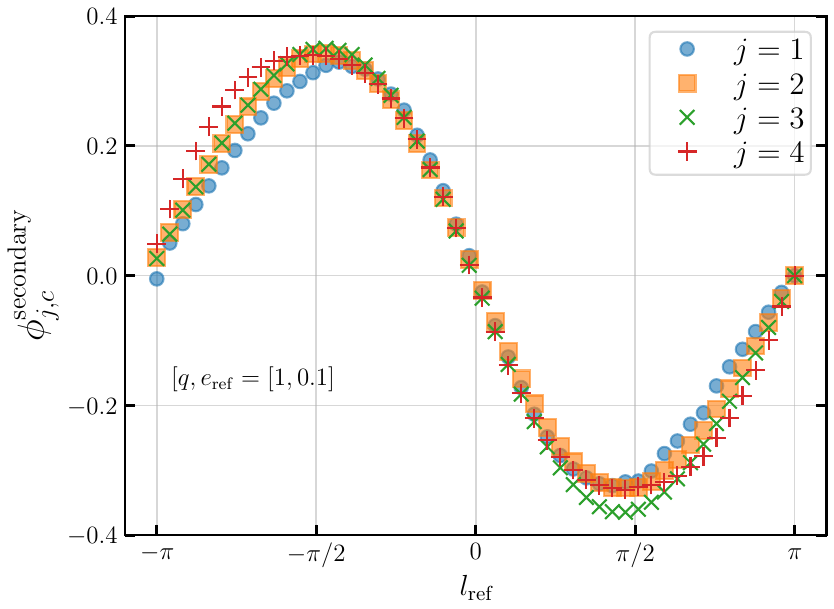}
\caption{We present the secondary phase corrections $\phi_{j,c}^{\rm secondary}$ for $[q, e_{\rm ref}=[1,0.1]$ for $j=[1,2,3,4]$. The plot suggests that these secondary phase corrections for the SVD coefficients are mostly independent of the value of $j$. Details are provided in Section~\ref{sec:model_mean_anomaly}.}
\label{fig:Cj_arg_secondary_same_for_different_j}
\end{figure}

\begin{figure}
\includegraphics[width=\columnwidth]{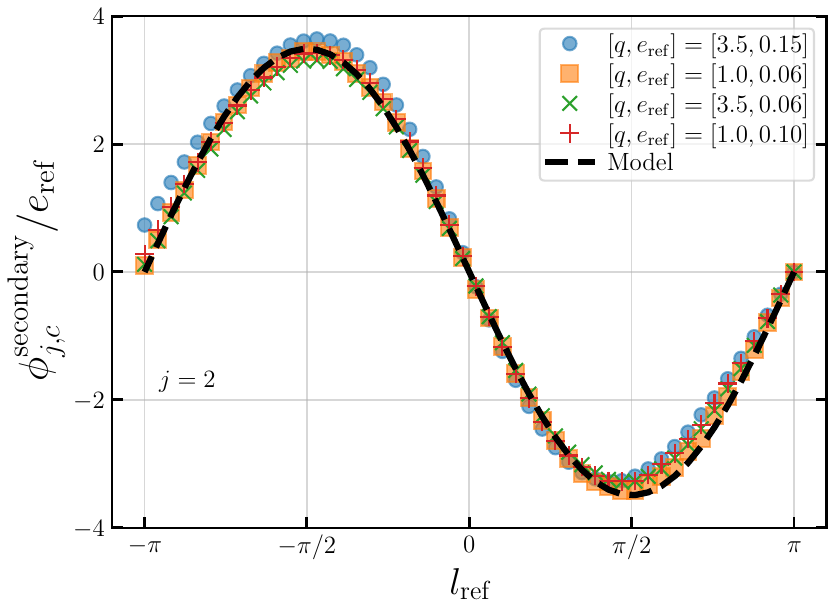}
\caption{We present the eccentricity-scaled (universal) secondary phase corrections $\phi_{j,c}^{\rm secondary}$ for $j=2$ for four different binary systems considered in Figure~\ref{fig:relative_Cj_vs_lref_combined}. Black dashed line indicate the best-fit model from Eq.(\ref{eq:secondary_phase_model}). Details are provided in Section~\ref{sec:model_mean_anomaly}.}
\label{fig:Cj_arg_ecc_scaled_secondary_phase}
\end{figure}

We now calculate the differences between the arguments of $C_{j}$ and the leading-order fits for each value of $j$. This is simply secondary phase corrections $\phi_{j,c}^{\rm secondary}$. We observe several interesting features. 
First, the secondary phase variations are more or less consistent across different harmonics. We illustrate this for $[q, e_{\rm ref}] = [1,0.1]$ in Figure~\ref{fig:Cj_arg_secondary_same_for_different_j}. 
Second, we find that these secondary features remain unchanged when varying the mass ratio $q$ while keeping the initial eccentricity $e_{\rm ref}$ fixed. 
Third, these features appear to be universal when analyzed in terms of an eccentricity-scaled quantity, defined as $\phi_{j,c}^{\rm secondary} / e_{\rm ref}$. We demonstrate these two features in Figure~\ref{fig:Cj_arg_ecc_scaled_secondary_phase} where we plot eccentricity-scaled secondary phase corrections $\phi_{j,c}^{\rm secondary}$ for $j=2$ for four different binary systems considered in Figure~\ref{fig:relative_Cj_vs_lref_combined}. We find that systems with different mass ratio values yield same eccentricity-scaled secondary phase $\phi_{j,c}^{\rm secondary} / e_{\rm ref}$. \textit{Overall, our results suggest that, at an approximate level, the eccentricity-scaled secondary phase contributions in the SVD coefficients exhibit universality.} Note that while we present our results for certain binary systems in this paper, we have verified these findings across a wide range of binaries with eccentricities up to $e_{\rm ref} \leq 0.2$ (the upper limit of NR calibration for the \texttt{TEOBResumS} model). 

Finally, we compute the mean of $\phi_{j,c}^{\rm secondary} / e_{\rm ref}$ for $j = [1,2,3,4]$ across a diverse set of binaries and obtain the following best-fit function:
\begin{equation}
\phi_{j,c}^{\rm secondary} = -3.491 \, e_{\rm ref} \sin(l_{\rm ref}).
\label{eq:secondary_phase_model}
\end{equation}
We show the best-fit function as dashed black line in Figure~\ref{fig:Cj_arg_ecc_scaled_secondary_phase}.

\subsection{Final modeling}
\label{sec:model}
Using the modeling insights from the previous two subsections (Sections~\ref{sec:model_svd_amp} and \ref{sec:model_svd_phase}), we find that if the eccentric harmonics are known at a reference mean anomaly (e.g., $l_{\rm ref} = \pi$), they can be directly obtained at other mean anomaly values for the same $q$ and $e_{\rm ref}$. Mathematically, the eccentric harmonics at $[q, e_{\rm ref}, l_{\rm ref} = \pi]$ are given by:  
\begin{equation}
h_{22,j,\rm ref}(t) = h_{22,j}(q,e_{\rm ref},l_{\rm ref} = \pi; t).
\end{equation}
The harmonics at any other mean anomaly can then be expressed as:  
\begin{equation}
h_{22,j}(q,e_{\rm ref},l_{\rm ref}; t) = h_{22,j,\rm ref}(t) A_{\rm corr} e^{i \phi_{\rm corr}},
\end{equation}
where $A_{\rm corr}$ is the amplitude correction and $\phi_{\rm corr}$ is the phase correction. We define:
\begin{equation}
A_{\rm corr} = 1 + \frac{\Delta |C_{j}|}{|C_{j,0}|}
\end{equation}
and 
\begin{equation}
\phi_{\rm corr} = {\rm Arg}[C_{j}] - {\rm Arg}[C_{j,0}].
\end{equation}
We obtain ${\rm Arg}[C_{j,0}]$ from our approximate model (described in Section~\ref{sec:model_svd_phase}) at $l_{\rm ref} = \pi$.

\begin{figure}
\includegraphics[width=\columnwidth]{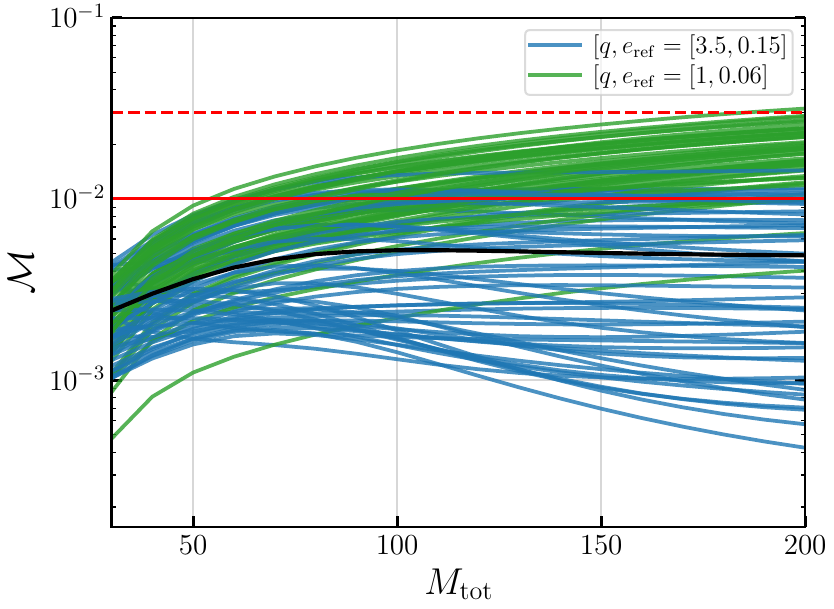}
\caption{We present the eccentricity-scaled (universal) secondary phase corrections $\phi_{j,c}^{\rm secondary}$ for $j=2$ for four different binary systems considered in Figure~\ref{fig:relative_Cj_vs_lref_combined}. Black dashed line indicate the best-fit model from Eq.(\ref{eq:secondary_phase_model}). Details are provided in Section~\ref{sec:model_mean_anomaly}.}
\label{fig:meanano_model_mismatches}
\end{figure}

\subsection{Modeling accuracy}
Next, we compute the frequency-domain mismatches (as defined in Ref.~\cite{Cutler:1994ys}) between actual \texttt{TEOBResumS} waveforms for different mean anomaly values and the approximate waveform using the modelling strategy in Section~\ref{sec:model}. We assume the Advanced LIGO sensitivity noise curve~\cite{KAGRA:2013rdx} for total masses varying from $M=20M_{\odot}$ to $M=200M_{\odot}$. We show the mismatches for two different binaries with $[q,e_{\rm ref}=[3.5,0.15]$ (blue curves) and $[q,e_{\rm ref}=[1,0.1]$ (orange curves). We find that in most cases, the mismatches remain below $10^{-2}$, The average mismatch is approximately $0.007$. However, for high-mass binaries (for which ring-down will dominate the signal), mismatches increase beyond $10^{-2}$ but do not exceed the $3\%$ mismatch threshold. This indicates that our model is reasonably good for detection and data analysis purposes. In our companion paper, we provide mismatches throughout the parameter space~\cite{Islam2025InPrep}.

%==========================================================================
%==========================================================================
%==========================================================================
\section{Summary and Future Work}
\label{sec:discussion}
To summarize, we have developed two data-driven complementary methods to extract eccentric harmonics from non-spinning eccentric waveforms. For simplicity, we showcase the efficiency using the quadrupolar mode of the waveform, but in principle, this framework will also work for any other spherical harmonic modes. Furthermore, there is no reason for this framework to fail in non-precessing binaries, as the waveform morphology does not change when aligned or anti-aligned spinning systems are considered instead of non-spinning systems (as studied in this paper).

Our methods take inspiration from Newtonian and PN calculations but are rooted in widely used data analysis techniques in signal processing and dimensionality reduction. The first method uses SVD to extract the harmonics from the basis vectors. The second method builds upon a low-pass filter around the expected frequencies of the eccentric harmonics (as known from Newtonian calculations). We find that the filter-based method is very fast but suffers from instability due to the Gibbs phenomenon at the beginning and end of the waveform time window. On the other hand, the SVD-based method is more stable than the filter-based method but is computationally expensive. Furthermore, the SVD-based method also suffers from the spilling of power from dominant basis vectors to others near the merger. To fix this, we designed simple smoothing techniques. Overall, we find that both methods yield the same eccentric harmonics and thereby validate each other.

Finally, we note that due to relativistic effects, the Newtonian expectations of the phase and frequency relations between different eccentric harmonic modes no longer hold. There are additional corrections that need to be taken into account. However, we find that these corrections follow a simple structure, which can be utilized in future modeling of these harmonics. On the other hand, we discover a simple way to model waveforms as a function of the mean anomaly, provided the harmonics are known at any reference mean anomaly value. This observation significantly simplifies the modeling of eccentric binaries, where capturing the mean anomaly dependence has traditionally been a challenge in semi-analytical or data-driven modeling.

Since these eccentric harmonics are monotonic functions of time and frequency, one can now develop relative binning or heterodyning methods for eccentric signal analysis~\cite{Zackay:2018qdy, Roulet:2024hwz} and build efficient template banks using the eccentric harmonics directly~\cite{Wadekar:2024zdq}—without relying on the more complicated full eccentric waveform, which exhibits non-monotonic features. This approach will significantly reduce the computational cost of data analysis and detection for eccentric systems. Overall, our work provides a new perspective on modeling eccentric binaries in GW science.

In the near future, we plan to extract these eccentric harmonics from the NR data. However, there are two main challenges. First, we cannot apply the \texttt{filter\_method}, because the Gibbs phenomenon would cause us to lose a significant portion of both the initial NR data and the merger–ringdown signal, rendering the effort ineffective. Second, the \texttt{svd\_method} requires an ensemble of waveforms that share the same eccentricity but differ in mean anomaly. Generating waveforms with identical eccentricity to high precision is challenging, although this has recently been explored in Ref.~\cite{Nee:2025zdy}. Therefore, in future work we will investigate whether a similar decomposition can be achieved using the NR data from the SXS collaboration, and whether such a decomposition can improve the efficiency of building NR surrogates.

%%%%%%%%%%%%%%%%%%%%%%%%%%%%%%%%%%%%%%%%%%%%%%%%%%%%%%%%%%%%%%%%%%%%%%%%%%%%%%%%%%%%%%%%%%%%%%%%%%%%%%%
%%%%%%%%%%%%%%%%%%%%%%%%%%%%%%%%%%%%%%%%%%%%%%%%%%%%%%%%%%%%%%%%%%%%%%%%%%%%%%%%%%%%%%%%%%%%%%%%%%%%%%%%
\begin{acknowledgments}
We thank Steve Fairhurst, Ben Patterson, Scott Field, Peter James Nee, Lucy Thomas and Antoni Ramos-Buades for useful discussions and comments.
This research was supported in part by the National Science Foundation under Grant No. NSF PHY-2309135 and the Simons Foundation (216179, LB). 
Use was made of computational facilities purchased with funds from the National Science Foundation (CNS-1725797) and administered by the Center for Scientific Computing (CSC). The CSC is supported by the California NanoSystems Institute and the Materials Research Science and Engineering Center (MRSEC; NSF DMR 2308708) at UC Santa Barbara. 
JR acknowledges support from the Sherman Fairchild Foundation. 
TV acknowledges support from NSF grants 2012086 and 2309360, the Alfred P. Sloan Foundation through grant number FG-2023-20470, the BSF through award number 2022136, and the Hellman Family Faculty Fellowship.
\end{acknowledgments}
%%%%%%%%%%%%%%%%%%%%%%%%%%%%%%%%%%%%%%%%%%%%%%%%%%%%%%%%%%%%%%%%%%%%%%%%%%%%%%%%%%%%%%%%%%%%%%%%%%%%%%%%
%%%%%%%%%%%%%%%%%%%%%%%%%%%%%%%%%%%%%%%%%%%%%%%%%%%%%%%%%%%%%%%%%%%%%%%%%%%%%%%%%%%%%%%%%%%%%%%%%%%%%%%%

%==========================================================================
%==========================================================================
%==========================================================================
%%%%%%%%%%%%%%%%%%%%%%%%%%%%%%%%%%%%%%%%%%%%%%%%%%%%%%%%%%%%%%%%%%%%%%%%%%%%%%%%%%%%%%%%%%%%%%%%%%%%%%%%
\bibliography{References}
%%%%%%%%%%%%%%%%%%%%%%%%%%%%%%%%%%%%%%%%%%%%%%%%%%%%%%%%%%%%%%%%%%%%%%%%%%%%%%%%%%%%%%%%%%%%%%%%%%%%%%%%

\appendix
\section{Comparison with Ref.~\cite{Patterson:2024vbo}}
\label{sec:patterson}
Recently, Ref.~\cite{Patterson:2024vbo} proposed a framework that combines waveforms with the same eccentricity but different mean anomaly (as obtained in Section~\ref{sec:svd_method})~\footnote{While Ref.~\cite{Patterson:2024vbo}  demonstrates their framework for aligned-spin binaries too, we restrict ourselves to non-spinning scenario for now.}:
\begin{equation}
h_k = \frac{1}{n} \sum_{i^\prime=0}^{n-1} e^{2\pi i i^\prime k} h_{i^\prime}
\end{equation}
with $k\in\{-1,0,1,2\}$.
Note that we choose to use the harmonic index $j=[1,2,3,4,\dots]$, as this convention is more common in Newtonian and PN literature. To compare, their $k=-1$ harmonic corresponds to our $j=1$ harmonic, and $k=[0,1,2]$ correspond to $j=[2,3,4]$, respectively.
To guarantee that the harmonics ${h_k}$ are mutually orthogonal, Ref.~\cite{Patterson:2024vbo} applies the Gram-Schmidt orthogonalization procedure. First, the harmonics are relabeled (with index $l=$ $\{0,1,2,3,4, \ldots\}$) in descending order of significance (i.e $l=$ $\{0,1,2,3,4, \ldots\}$ correspond to harmonics with $k=$ $\{0,1,-1,2,3, \ldots\}$). The relabeled harmonics are then orthogonalized as:
\begin{equation} 
h_l=h_l-\sum_{i=0}^{l-1} \frac{\left(h_l \mid h_i\right)}{\left(h_i \mid h_i\right)} h_i. 
\end{equation} 

Their framework does not guarantee that the eccentric harmonics obtained will vanish when the eccentricity approaches zero. Instead, due to the manner in which these harmonics are constructed, there are possibilities of mode mixing, leading to merger-ringdown contributions from $j=2$ spilling into higher eccentric harmonics. Furthermore, there is no explicit attempt to ensure that the eccentric harmonics are smooth, monotonic functions of time and frequency. Such non-monotonicity is observed in Ref.~\cite[figure~4]{Patterson:2024vbo}. Overall, while the approach presented in Ref.~\cite{Patterson:2024vbo} provides harmonics which are close to the eccentric harmonics extracted in this paper, they will not be smooth harmonics (e.g. Fig.~\ref{fig:ecc_harm_hierarchy} in this paper) that becomes negligible around merger as opposed to the ones we provide.

\begin{figure}
\includegraphics[width=\columnwidth]{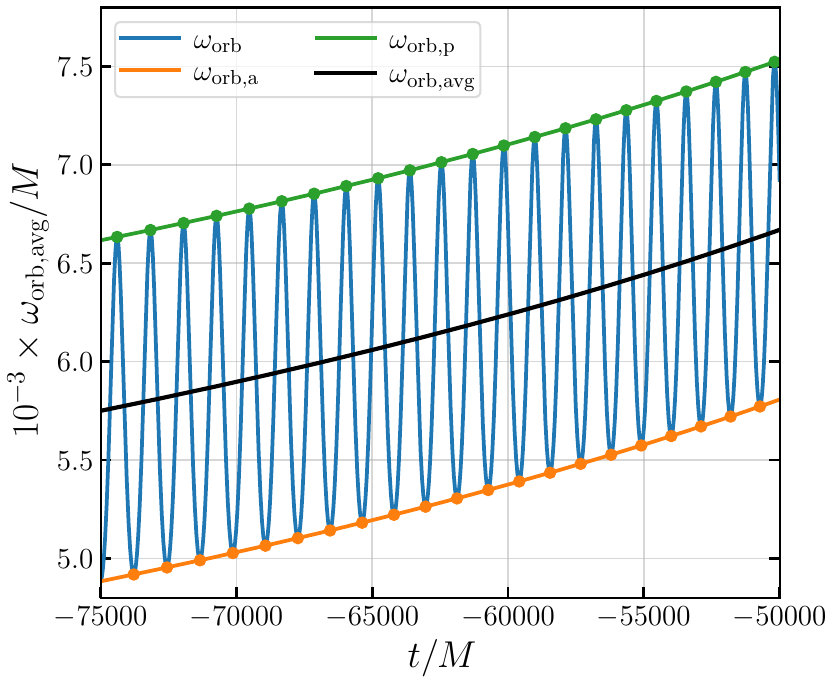}
\caption{We show the instantaneous orbital frequency $\omega_{\rm orb}$ (blue line) of the full eccentric quadrupolar waveform shown in Fig.~\ref{fig:spectogram} for the first $25000M$ of the time window for visual clarity. We point out that while $\omega_{\rm orb}$ is oscillatory, it is possible to identify the envelopes corresponding to the periastron ($\{\omega_{\rm orb, \rm p}\}$; green line) and apastron ($\{\omega_{\rm orb, \rm a}\}$; orange line) and obtain a monotonic average instantaneous orbital frequency $\omega_{\rm orb, \rm avg}$ (black line). Details are in Section~\ref{sec:avg_orb_phase}.
}
\end{figure}

\begin{figure}
\includegraphics[width=\columnwidth]{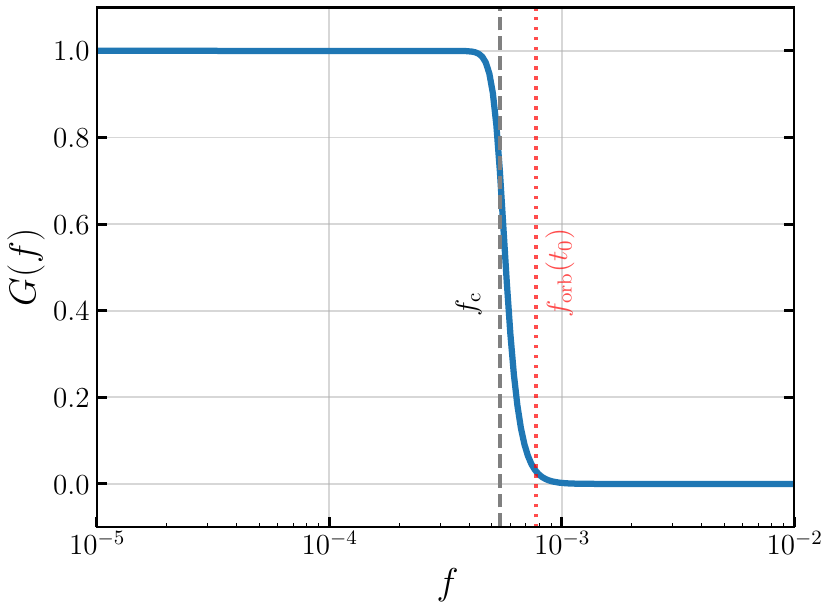}
\caption{We show the frequency response $G(f)$ of the Butterworth low-pass filter used to obtain eccentric harmonics from the full quadrupolar waveform shown in Fig.~\ref{fig:spectogram}. The dashed red vertical line indicates the initial orbital frequency of the signal, while the dashed grey line denotes the threshold frequency used to create the low-pass filter with order 10. Details are provided in Section~\ref{sec:lpfilter}.}
\label{fig:filter_frequency_response}
\end{figure}

On top of that, the harmonics presented in Ref.~\cite{Patterson:2024vbo} are not fully the eccentric harmonics as these harmonics retain small but non-zero SNR in the circular limit (Ref.~\cite{Patterson:2024vbo}, Figure 5). On the other hand, eccentric harmonics presented in this work will go to zero by construction and have zero SNR in the circular limit.

%==========================================================================
%==========================================================================
%==========================================================================
\section{Calculating the average orbital phase from quadrupolar mode}
\label{sec:avg_orb_phase}
We use the quadrupolar mode of the radiation $h_{22}(t)$ to calculate the instantaneous frequency $\omega_{22}(t)$. This immediately gives us the instantaneous orbital frequency:
\begin{align}
\omega_{\rm orb}(t) = \frac{\omega_{22}(t)}{2}.
\end{align}

However, due to eccentricity, the instantaneous orbital frequency exhibits oscillatory features related to the periastron and apastron of an elliptical orbit. To remove this oscillation, we calculate a continuous representation of $\omega_{\rm orb}$ at the periastron and apastron points. To do this, we first identify the times $\{t_p\}$ and instantaneous orbital frequency $\{\omega_{\rm orb, \rm p}\}$ corresponding to the maxima in the original $\omega_{\rm orb}(t)$ time series. We perform the same procedure for the minima, which gives us the relevant quantity for the apastron. We then include the instantaneous orbital frequency at the merger time to both $\{\omega_{\rm orb, \rm p}\}$ and $\{\omega_{\rm orb, \rm a}\}$. 
\begin{figure}
\includegraphics[width=\columnwidth]{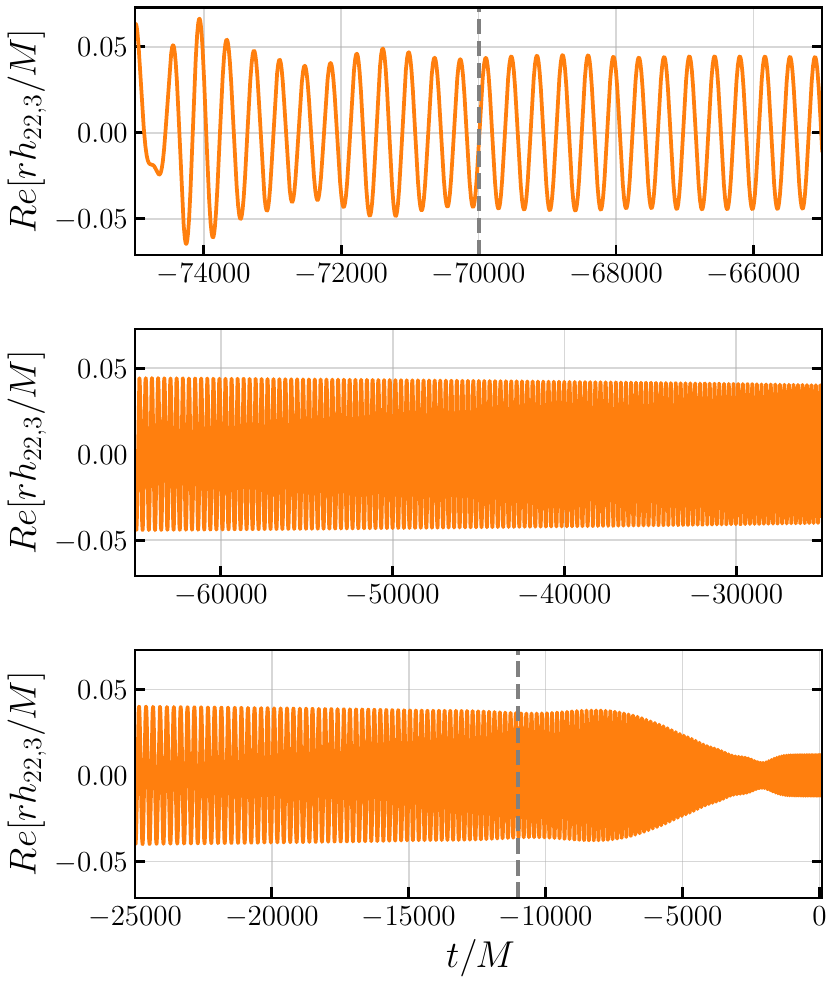}
\caption{We show noises due to Gibbs phenomenon at the start and end of the $j=3$ eccentric harmonic extracted using the filter-based method from the quadrupolar mode shown in Fig.~\ref{fig:spectogram}. Dashed vertical line show the start and end of the clean harmonic data. Details are provided in Section~\ref{sec:filter_limitation}.}
\label{fig:filter_j3}
\end{figure}

The continuous representations are obtained using cubic splines from \texttt{scipy.interpolate}~\footnote{\href{https://docs.scipy.org/doc/scipy/reference/interpolate.html}{https://docs.scipy.org/doc/scipy/reference/interpolate.html}}. Once we have continuous representations of $\omega_{\rm orb}$ at the periastron and apastron points, we calculate the average instantaneous orbital frequency: $\omega_{\rm orb, \rm avg}(t)$.
Details of such frameworks have previously appeared in the literature, and our prescription closely follows those. Finally, we integrate $\omega_{\rm orb, \rm avg}(t)$ to obtain the average orbital phase. For simplicity, we set the integration constant to zero:
\begin{align}
\phi_{\rm orb, \rm avg}(t) = \int \omega_{\rm orb, \rm avg}(t) \, dt.
\end{align}

%==========================================================================
\section{Constructing a low-pass filter}
\label{sec:lpfilter}
The impulse response $g(t)$ is the inverse Fourier transform of the frequency response $G(f)$. The Butterworth low-pass filter has a frequency response:
\begin{equation}
    G(f) = \frac{1}{\sqrt{1 + \left(\frac{f}{f_c}\right)^{2N}}}
\end{equation}
where $f_c$ is the cutoff frequency and $N$ is the filter order. For our purposes, we use $N = 10$.

%==========================================================================
\section{Limitations of the filter-based method}
\label{sec:filter_limitation}
While \texttt{filter\_method} provides clean eccentric harmonics for most of the time, it fails to provide any meaningful estimate of the harmonics around the merger and at the very early times due to noises originated from Gibbs phenomenon. However, at very early times, waveforms as well as the harmonics change very slowly, and it is possible to extrapolate the behavior. But at late times around merger, which contains most of the power in BBH mergers, we have no known way to properly reconstruct the harmonics. We show this noisy features for $j=3$ eccentric harmonic in Fig.~\ref{fig:filter_j3} where we lose $\approx5000M$ at the start of the waveform and $\approx 10000M$ at the end. Similar loss happens in other eccentric harmonics too.

\end{document}